\documentclass[prb,twocolumn, floatfix,nofootinbib]{revtex4-2}
\usepackage{physics}
\usepackage{bm}
\usepackage[]{amsmath}
\usepackage{amssymb}
\usepackage[]{color}
\usepackage[dvipsnames]{xcolor}
\usepackage[]{graphicx}
\usepackage{tabularx}
\usepackage{comment}
\usepackage{subfigure}
\usepackage{chemformula}
\usepackage{siunitx}
\renewcommand\vec{\boldsymbol}

\DeclareMathOperator{\sgn}{sgn}
\DeclareMathOperator{\re}{Re}
\DeclareMathOperator{\im}{Im}
\newcommand{\iu}{\mathrm{i}} 
\newcommand{\eu}{\mathrm{e}} 

\newcommand{\be}{\begin{equation}}
\newcommand{\ee}{\end{equation}}

\newcommand*\diff{\mathop{}\!\mathrm{d}}
\newcommand{\non}{\nonumber}
\newcommand{\dg}{\dagger}
\newcommand{\ve}[1]{{\boldsymbol #1}}
\newcommand{\qv}{\ve{q}}
\newcommand{\kv}{\ve{k}}

\newcommand{\veM}{{\bm{M}}}
\newcommand{\vem}{{\bf{m}}}

\newcommand{\Dh}{{\rm D}_{2h}}
\newcommand{\AVS}{AV$_3$Sb$_5$}
\newcommand{\KVS}{KV$_3$Sb$_5$}
\newcommand{\CVS}{CsV$_3$Sb$_5$}
\newcommand{\RVS}{RbV$_3$Sb$_5$}
\def\em{\it}

\allowdisplaybreaks

\begin{document}

\title{Electronic instabilities of kagom\'e metals: saddle points and Landau theory}

\author{Takamori Park}
\affiliation{Department of Physics, University of California, Santa
Barbara, CA 93106-9530, USA}
\author{Mengxing Ye}
\affiliation{Kavli Institute for Theoretical Physics, University of
  California, Santa Barbara, CA 93106-4030, USA}
\author{Leon Balents}
\affiliation{Kavli Institute for Theoretical Physics, University of
  California, Santa Barbara, CA 93106-4030, USA}
\affiliation{Canadian Institute for Advanced Research, 661
University Ave., Toronto, ON M5G 1M1 Canada}

\date{\today}

\begin{abstract}
    We study electronic instabilities of a kagom\'e metal with a Fermi energy close to saddle points at the hexagonal Brillouin zone face centers. Using parquet renormalization group, we determine the leading and subleading instabilities, finding superconducting, charge, orbital moment, and spin density waves.  We then derive and use Landau theory to discuss how different primary density wave orders give rise to charge density wave modulations, as seen in the \AVS\ family, with A=K,Rb,Cs.  The results provide strong constraints on the mechanism of charge ordering and how it can be further refined from existing and future experiments.
\end{abstract}
\maketitle

\section{Introduction}
Two dimensional correlated metals based on transition metal ions are a
classic subject for many body physics~\cite{MITRMP1998,CorrelateMetalReview2014}, displaying diverse electronic
phenomena such as unconventional superconductivity~\cite{IronSCreview2010,SCReview2012,sato2017topological}, charge and spin
order~\cite{TsenTaS22015,ye2018massive}, nematicity~\cite{Fisher2011Nematicity,SatoNematicity2017}, strange metallic behavior~\cite{VarmaStangeMetal1989,NFLRMP2001,StrangeMetalRMP2003}, and more. These topics have been most heavily investigated in theory and experiment in
structures based on square lattices in the cuprates~\cite{SCReview2012,VarmaSCReview} and Fe
superconductors~\cite{IronSCreview2010,HighTcIron2012,HighTcIron2016}.  Correlated metals with hexagonal/triangular symmetry
 are much less common, with the best-known example being the
triangular lattice cobaltates~\cite{TakadaCobalt2003,OngCobalt2004}, which however are complicated by Na vacancy ordering and water intercalation\cite{PhysRevB.68.214517}.

Recently, a new class of kagom\'e metals, with chemical formula \AVS,
where A\;=\;K, Rb, or Cs, have emerged as an exciting realization of
quasi-2D correlated metals with hexagonal symmetry~\cite{ortiz2019new}.  These materials have
been shown to display several electronic orders setting in through
thermodynamic phase transitions: multi-component (``3Q'') hexagonal
charge density wave (CDW) order below a  $T_c \approx 90K$~\cite{jiang2020discovery,ZeljkovicCascade2021,2021arXiv210309769L,UykurOptical2021,2021arXiv210304760L,ZhouKagomeOptical2021,kenney21:_absen_kv3sb,ortiz2021fermi}, and superconductivity with critical temperature of $2.5K$ or smaller~\cite{BrendenSC2021,jiang2020discovery,ZeljkovicCascade2021,Zhao2102KagomeSC,Chen2102KagomeSC,Chen2103KagomeSC,Duan2103KagomeSC,Zhang2103KagomeSC,Mu2104KagomeSC,Ni2014KagomeSC}, and some indications of nematicity and
one-dimensional charge order in the normal and superconducting states~\cite{ZeljkovicCascade2021,Chen2103KagomeSC,xiang2021nematic}.
Other experiments show a strong anomalous Hall effect~\cite{wang2020giant,Yu2021AHE}, suggesting
possible topological physics.  Furthermore, density functional calculations identified \CVS\ as a $\mathbb{Z}_2$ topological metal~\cite{BrendenZ22020}. Angle resolved photoemission studies show these materials to be multi-band systems with several Fermi surface components~\cite{BrendenZ22020}, including
approximately nested components and a Fermi energy that
is close to multiple saddle points of the dispersion, in agreement
with density functional theory~\cite{BrendenZ22020,Zhao2103DFT}. Furthermore, strong momentum dependent charge gaps near the saddle point momenta were observed below the CDW transition temperature~\cite{Liu2104photo,Wang2104photo}.    

Many of these ingredients bring to mind a storied idea of electronic
instabilities enhanced by van Hove singularities near saddle points.
This mechanism figured heavily in early theoretical treatments of the
cuprates~\cite[and references therein]{PhysRevLett.56.2732,MarkiewiczVHS1997,HurVHS2009}.  In a two-dimensional system, the divergence of the density
of states generates enhanced scattering amongst electrons near the
saddle points of the bands, which may drive not only superconducting
but also other charge and spin instabilities.  The same idea emerged
recently in the context of doped graphene~\cite{nandkishore2012chiral,nandkishore_itinerant_2012,nandkishore2012interplay}, and has been applied to
the theory of magic angle graphene bilayers~\cite{LinYP2019,Chichinadze2020a,Chichinadze2020b,Classen2020,LinYP2020}.  The observations in \AVS\ suggest another application.  Notably, the observed period of charge
order in \AVS\ within the two-dimensional kagom\'e plane is precisely
that expected from scattering amongst the saddle points located at the hexagonal Brillouin zone face centers (denoted as $\veM$).  

While the saddle point model on a 2D hexagonal lattice has been studied extensively in the literature~\cite{nandkishore2012chiral,nandkishore2012interplay,nandkishore_itinerant_2012,LinYP2019}, its application to a coherent understanding of the electronic instabilities revealed in multiple experiments remains to be understood. First, the renormalization group studies from repulsive interactions~\cite{nandkishore2012chiral,LinYP2019} suggest leading spin density wave, superconductivity, and orbital moment instabilities, rather than the charge density wave order which is observed.  Therefore it is natural to expect that the lattice plays some role, and thus it is important to understand the combined effect of electron-phonon  and electron correlation.  Second, the strong anomalous Hall effect observed below the charge density wave critical temperature suggests time-reversal symmetry breaking order may develop below $T_c$. It would be desirable to understand the interplay between time-reversal symmetric and broken CDW order. Third, in addition to the $2\times 2$ charge density wave within the 2D layer, STM and x-ray studies also observed modulation in the z-direction, i.e.\ $k_z\neq 0$, in \RVS\ and \CVS. The three-dimensional alignment of 2D charge density waves has not yet been studied theoretically. 

In this paper, we explore the electronic instabilities due to
interactions amongst electrons near the saddle points with hexagonal
symmetry.  We apply the parquet renormalization group scheme~\cite{ZheleznyakPRG1997,MetznerPRG1998,ChubukovPRG2008,ChubukovPRG2009} to
determine all the primary and secondary instabilities, generalizing
prior work~\cite{nandkishore2012chiral,LinYP2019}. We further use mean field theory to study different possible emergent density wave states, which include not only a
conventional charge density wave, but also spin and orbital moment
density waves.  We investigate how they relate to the observed charge
density wave measured in \AVS.  Through this analysis, we provide
constraints on the interpretation of experimental observations, and
suggestions for future theoretical and experimental studies.  

The rest of this paper is structured as follows: In Sec.~\ref{sec:Model}, we construct a low energy continuum model by taking patches around the saddle point momenta $\veM$ and identify its connection to the real space tight-binding model. In Sec.~\ref{sec:RG}, we generalize the parquet renormalization group formulation, and discuss all stable fixed points within the patch model. In Sec.~\ref{sec:MFT}, we analyze the mean field theory for each fixed point solutions. This includes a Landau theory analysis of the conventional charge density wave (Sec.~\ref{sec:RCDWMFT}) as well as an analysis of other leading instabilities, i.e.\ the orbital moment density wave (Sec.~\ref{sec:ICDWRCDWMFT}) and spin density wave (Sec.~\ref{sec:RSDWRCDWMFT}), and their interplay with the conventional charge density wave. In Sec.~\ref{sec:extens-exper-impl}, we discuss the implications of the results obtained in Sec.~\ref{sec:MFT} to experimental observations. The real space pattern on the kagom\'e lattice are shown in Sec.~\ref{sec:realspaceCDW}. To understand the three dimensional 3Q CDW order with $k_z\neq 0$, we study the effects of weak interlayer coupling and discuss the allowed $k_z$ for both conventional CDW and orbital moment density wave orders. The experimentally observed anisotropic CDW order can also be explained within this picture. In Sec.~\ref{sec:critical-behavior}, we discuss the critical behavior for different types of charge order instabilities. The staggered and uniform orbital moment is estimated in Sec.~\ref{sec:magn-moment-induc}. The possibility of a magnetic field induced mixture of conventional CDW and orbital moment is also briefly discussed there. Lastly, a summary of our work and discussions on the theoretical and experimental implications are presented in Sec.~\ref{sec:summary}. 

\section{Model}\label{sec:Model}
\begin{figure}[htp]
  \subfigure[]{\includegraphics[width=0.49\linewidth]{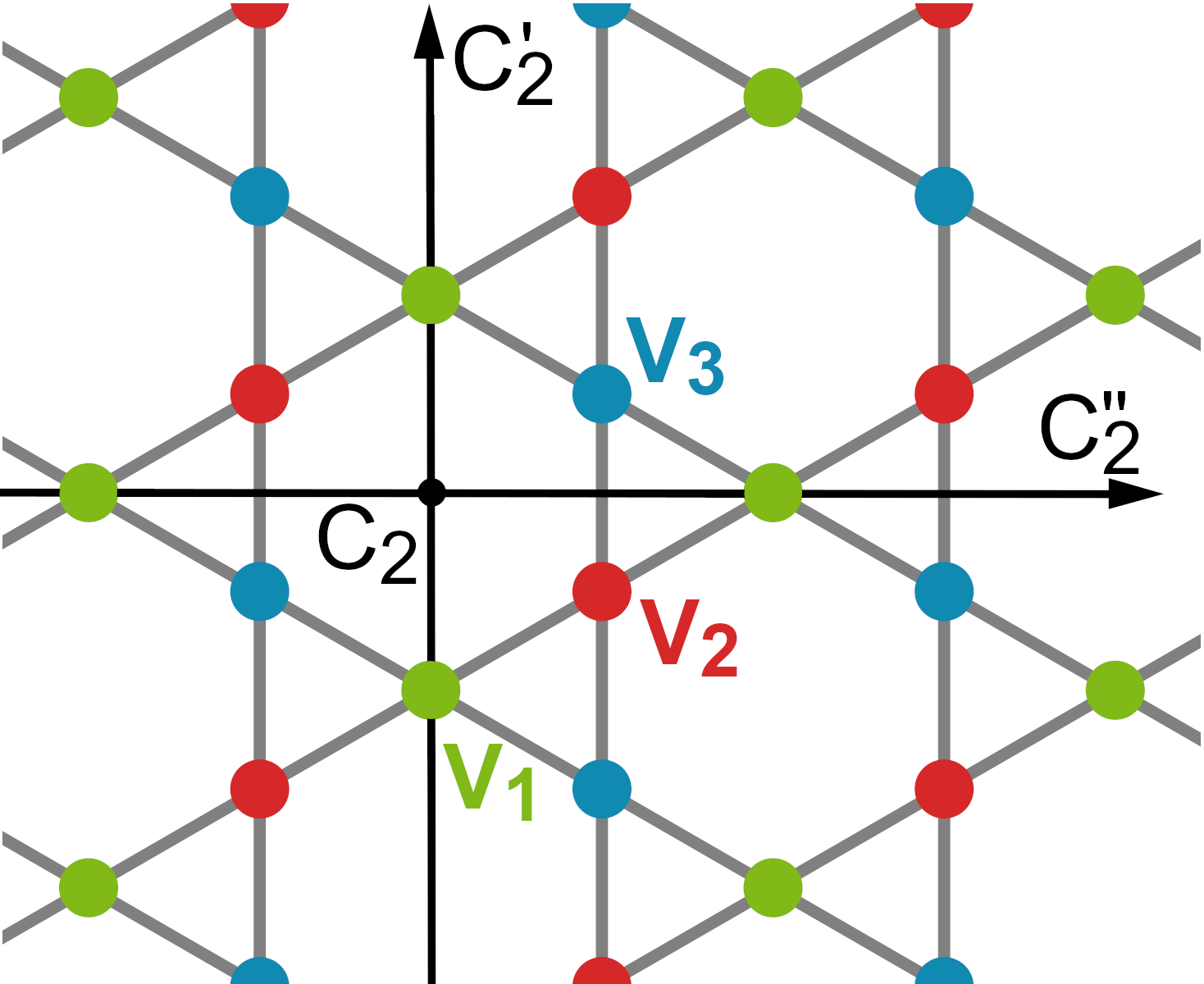}\label{fig:lattice}}
  \subfigure[]{\includegraphics[width=0.49\linewidth]{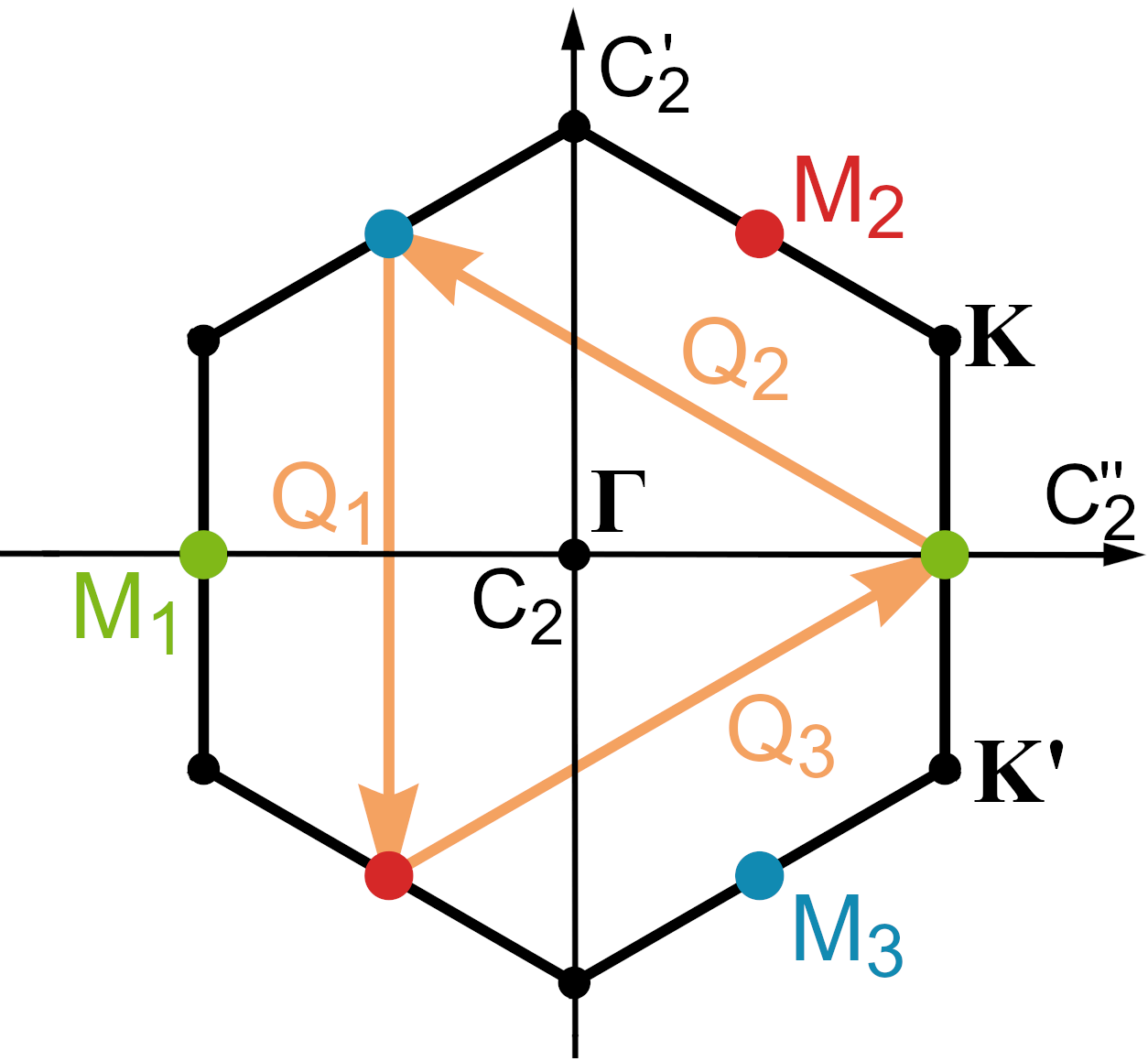}\label{fig:BZ}}
  \caption{\subref{fig:lattice}Vanadium kagom\'e lattice structure in \AVS. The actual lattice structure has
  a Sb atom in the center of the hexagonal faces, but they are not shown here since our work is only focused on the electronic
  behavior of vanadium. The space group of the full lattice is $P6/mmm$.
  \subref{fig:BZ} Hexagonal Brillouin zone of \AVS. There are three distinct M points along the Brillouin
  zone boundaries. The $\veM_\alpha$ points in the Brillouin zone are connected by three nesting vectors $\{\vec{Q}_\alpha\}$. The
  nesting vectors satisfy $\vec{Q}_\alpha\equiv-\vec{Q}_\alpha$ up to a reciprocal lattice vector.} \label{fig:latticeBZ}
\end{figure}

\begin{figure}[htp]
  \begin{center}
    \includegraphics[width=\linewidth]{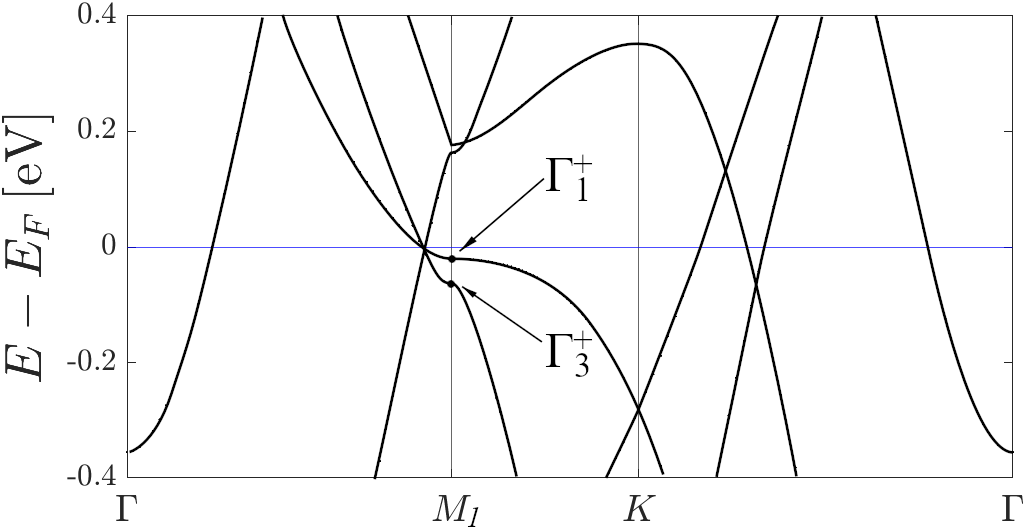}
  \end{center}
  \caption{Band structure of \CVS\ calculated using DFT without spin-orbit coupling \cite{Teicher}. The Fermi energy is set to \SI{6.81}{\eV}.
  $\Gamma_1^+,\Gamma_3^+$ are the irreducible representations of the two saddle point bands nearest the Fermi energy at
  $\veM_1$. The little co-group at $\veM_1$ is $D_{2h}$, and the orientation of the symmetry axes $C_2,C_2',C_2''$ are shown
  in Fig.~\ref{fig:latticeBZ}. We use the notation found in Ref. \cite{koster_properties}.
  \label{fig:DFT}}
\end{figure}

In this section, we introduce a simple model to study the electronic behavior in \AVS. As explained in the
introduction, STM and x-ray scattering measurements show a $2\times2$ CDW order with wavevector $\bm{Q}_\alpha=\bm{Q}_\textrm{Bragg}/2$~\cite{ZeljkovicCascade2021, jiang2020discovery, 2021arXiv210304760L, 2021arXiv210309769L}. 
These wavevectors are equivalent to the momenta that connect the three $\ve{M}_\alpha$ points in the hexagonal Brillouin zone as
shown in Fig.~\ref{fig:latticeBZ}(b). In addition, DFT calculations show that the band structure has saddle points at the
$\veM_\alpha$ points near the Fermi level as shown in Fig.~\ref{fig:DFT} for \CVS. In two-dimensional systems, a saddle point is a van
Hove singularity with a logarithmically diverging density of states.  Based on these two observations, we assume that
the collective electronic behavior in \AVS\ is determined by the saddle points located at the $\ve{M}_\alpha$ points and the
interactions between them.

From these considerations, we construct a non-interacting low-energy continuum model by taking patches around the $\veM_\alpha$
points in the Brillouin zone with cutoff radius $\Lambda$. The Hamiltonian of such a model is
\begin{equation}
  H_0 =\sum_{\alpha}\sum_{\abs{\vec{q}}<\Lambda}c^\dagger_{\alpha
  \vec{q}}\left[\varepsilon_\alpha(\vec{q})-\mu\right]c_{\alpha \vec{q}},\\
  \label{eq:noninteractinghamiltonian}
\end{equation}
where $\alpha, \beta$ are patch indices, $\vec{q}$ is the momentum measured from $\veM_\alpha$,
$\varepsilon_\alpha(\vec{q})$ is the saddle point dispersion sitting at $\veM_\alpha$, and $\mu$ is the chemical potential measured away from the saddle point. Up to quadratic order in $\vec{q}$, the saddle point dispersions take the general
form,
\begin{align}
  \varepsilon_1(\vec{q})=&a q_x^2-bq_y^2,\notag\\
  \varepsilon_2(\vec{q})=&\frac{a-3b}{4} q_x^2+\frac{\sqrt{3}(a+b)}{2}q_xq_y+\frac{3a-b}{4}q_y^2,\notag\\
  \varepsilon_3(\vec{q})=&\frac{a-3b}{4} q_x^2-\frac{\sqrt{3}(a+b)}{2}q_xq_y+\frac{3a-b}{4}q_y^2,
  \label{eq:SPdispersion}
\end{align}
where $\vec{q}=(q_x,q_y)$.
The parameters $a,b$ that determine the shape of the saddle point must have the same sign, $ab>0$. Note, that the
dispersion of the three patches are related by a three-fold rotation. The condition for perfect nesting is given by
$a/b=3$, but in our work we will not necessarily impose this condition unless otherwise stated.

In general, there are four different cases of the form of the
dispersion that we can consider: (i) $a,b>0$, $a/b>1$, (ii) $a,b>0$,
$a/b<1$, (iii) $a,b<0$, $a/b>1$, and (iv) $a,b<0$, $a/b<1$.  These
cases can be all be mapped to case (i) by using two transformations.
First, rotating the coordinates by $\pi/2$ takes cases (iii), (iv) to
cases (ii), (i) respectively.  Second, a particle-hole
transformation combined with the same $\pi/2$ rotation maps case (ii)
to case (i) (changing the sign of $\mu$).  Hence, we only need to
consider case (i).  Note that the latter particle-hole-like
transformation becomes a symmetry when $a=b$ and $\mu=0$.

\begin{figure}[htp]
  \begin{center}
    \includegraphics[width=0.65\linewidth]{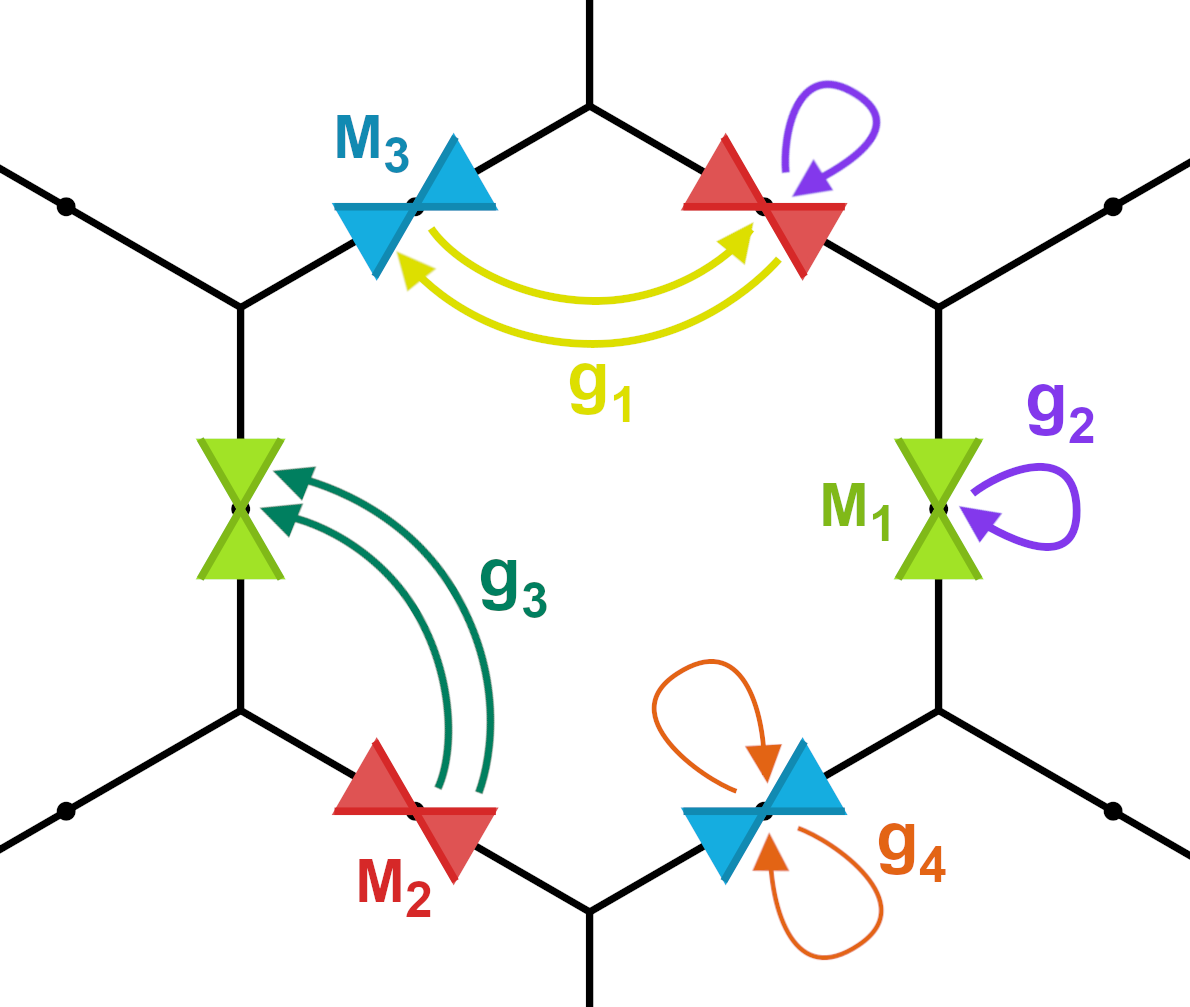}
  \end{center}
  \caption{All possible interactions in the patch model. The cones represent the saddle point dispersion at each $\veM_\alpha$ and
  the arrows denote the scattering processes described by the interactions.}
  \label{fig:interactions}
\end{figure}

Next, we introduce interactions to the continuum model by listing all possible electron-electron interactions between
the fermions in the three patches. There are four such interactions,
\begin{align}
  H_1
  =&\frac{1}{2\mathcal{N}}\sideset{}{'}\sum_{\abs{\vec{q}_1},\cdots,\abs{\vec{q}_4}<\Lambda}
  \bigg[\sum_{\alpha\neq\beta}\big(g_1
  c^\dagger_{\alpha \vec{q}_1 \sigma}c^\dagger_{\beta \vec{q}_2 \sigma'}
  c_{\alpha\vec{q}_3 \sigma'}c_{\beta \vec{q}_4 \sigma}\notag\\
  &\qquad+g_2c^\dagger_{\alpha \vec{q}_1\sigma}c^\dagger_{\beta \vec{q}_2\sigma'}c_{\beta \vec{q}_3\sigma'}c_{\alpha
  \vec{q}_4\sigma}\notag\\
  &\qquad+g_3c^\dagger_{\alpha \vec{q}_1\sigma}c^\dagger_{\alpha \vec{q}_2\sigma'}c_{\beta \vec{q}_3\sigma'}c_{\beta \vec{q}_4\sigma}\big)\notag\\
  &\qquad+\sum_\alpha g_4c^\dagger_{\alpha \vec{q}_1\sigma}c^\dagger_{\alpha \vec{q}_2\sigma'}
  c_{\alpha \vec{q}_3\sigma'}c_{\alpha \vec{q}_4\sigma}\bigg],
  \label{eq:interaction}
\end{align}
where
$\sideset{}{'}\sum_{\abs{\vec{q}_1},\cdots,\abs{\vec{q}_4}<\Lambda}\equiv\sum_{\abs{\vec{q}_1},\cdots,\abs{\vec{q}_4}<\Lambda}
\delta_{\vec{q}_1+\vec{q}_2,\vec{q}_3+\vec{q}_4}$, $\mathcal{N}$ is the number of unit cells in the system, and $g_i$ are the
interactions that are defined to be intrinsic and have units of energy.  As seen in Fig.~\ref{fig:interactions}, the
$g_1, g_2, g_3, g_4$ interactions represent inter-patch exchange, inter-patch density-density, Umklapp, and intra-patch
density-density scattering processes, respectively.

The patch model introduced here has been used in previous works to study the interaction and competition between different
instabilities in the Hubbard model and doped graphene \cite{furukawa_truncation_1998, nandkishore2012chiral}. An important thing to note is that the continuum model
defined above does not include details on the underlying structure of the lattice and the global band structure. We
supplement the results of our model with the symmetry information provided by the DFT results for \CVS\ shown in Fig.~\ref{fig:DFT} \cite{Teicher}. This gives a one-to-one correspondence between Brillouin zone patches, $\veM_\alpha$, and the vanadium sites, $V_\alpha$ (see App.~\ref{app:irrep}). A minimal tight binding model on the kagom\'e lattice can be readily obtained from this correspondence to faithfully describe the saddle point fermions (see Sec.~\ref{sec:icdwpattern} for more discussions).

\section{Renormalization group analysis}
\label{sec:RG}
\begin{figure}[htp]
    \centering
    \includegraphics[width=0.9\columnwidth]{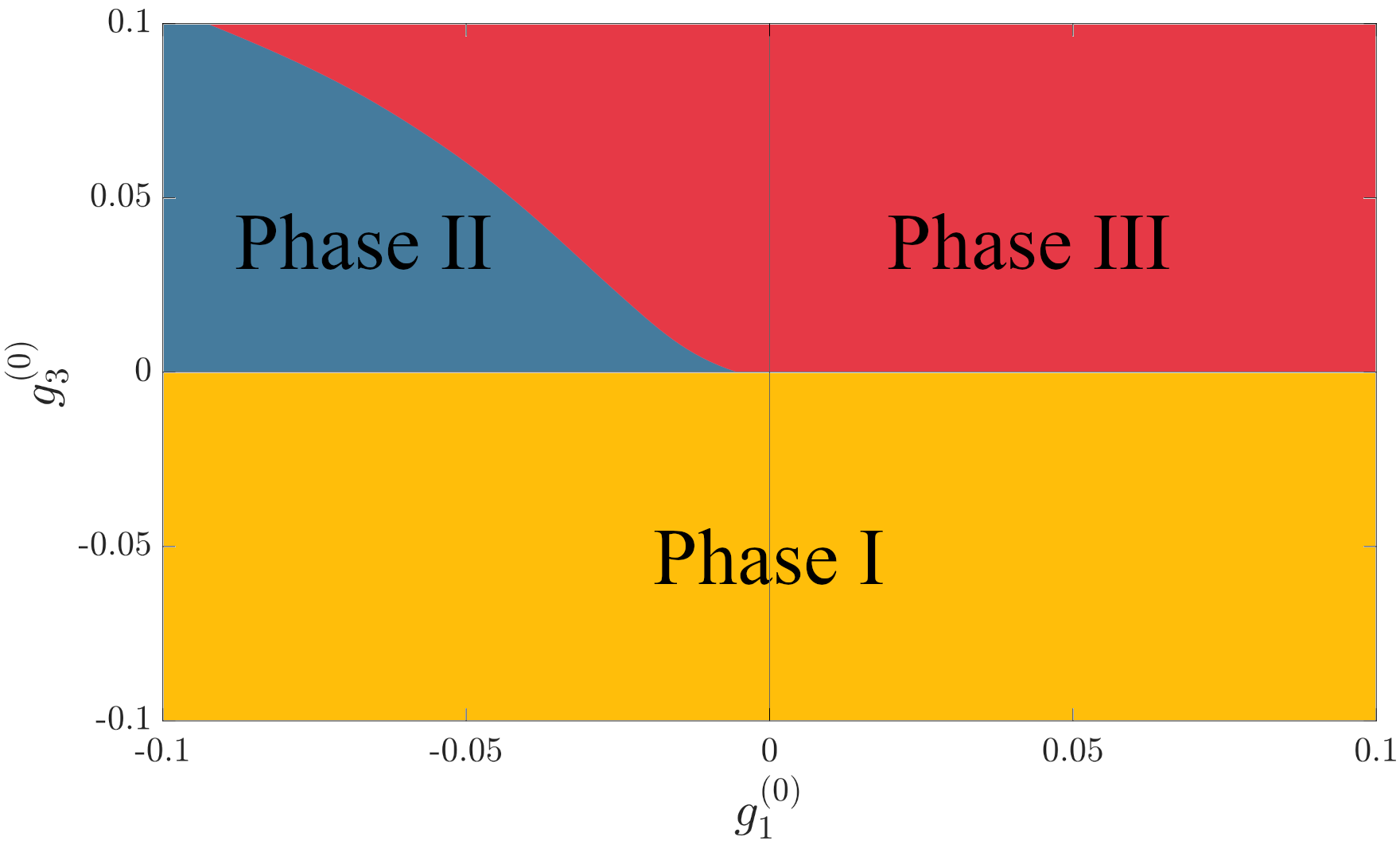}
    \caption{RG phase diagram for $g_2^{(0)}>0$. The leading instabilities of each phase under RG are, Phase I: sSC,
    rCDW, iSDW, Phase II: iCDW and rCDW, and Phase III: dSC, rSDW, iCDW.  The phase diagram was calculated assuming
    $d_1=1/2\,,g_2^{(0)}=0.1\,,g_4^{(0)}=0.3$.} 
    \label{fig:RG}
\end{figure}
While all four interactions $g_1, g_2, g_3$ and $g_4$ are marginal
at tree level, they acquire leading double logarithmic corrections
from particle-particle fluctuations at zero momentum ($\Pi_{\rm pp}
(0)$) and particle-hole fluctuations at momentum transfer
$\veM_\alpha$ ($\Pi_{\rm ph}(\veM_\alpha)$) at the one loop level and so
can become marginally relevant. To study the evolution of $g_i$ at
energy $E$ as the fermion fluctuations from cutoff energy $\Lambda_{\rm RG}\sim t \Lambda^2$
to $E$ are integrated out, we perform a parquet renormalization group
(pRG) analysis. Following Ref.~\cite{nandkishore2012chiral}, the pRG equations  for $g_i$ are
\begin{align}\label{eq:RGflow}
\frac{\diff g_1}{\diff y} &= 2d_1 g_1(g_2-g_1),\non\\
\frac{\diff g_2}{\diff y} &= d_1(g_2^2+g_3^2),\non\\
\frac{\diff g_3}{\diff y} &= -g_3^2-2 g_3 g_4 + 2 d_1 g_3 (2 g_2-g_1),\non\\
\frac{\diff g_4}{\diff y} &= - 2g_3^2 - g_4^2,
\end{align}
where $y= \Pi_{\rm pp}(0,E)\sim \ln^2 (\Lambda_{\rm RG}/E)$ is the RG time and
$d_1(y)=\diff \Pi_{\rm ph}(M_a) /\diff y$ is the ``nesting parameter",
which satisfies $0<d_1(y)\leqslant 1/2$.
For perfect nesting, $d_1(y)=1/2$ and is independent of RG time.
For concreteness, we will consider the perfect nesting case
hereafter, i.e.\ $d_1\equiv 1/2$. It has been checked that other
$0<d_1(y_c)<1/2$ does not qualitatively change the fixed point solutions and leading
density wave instabilities.

The RG equations, having entirely quadratic $\beta$-functions (the
right hand sides of the pRG equations), do not
have any non-trivial controlled fixed points in the usual sense.
Rather, they describe flows in the vicinity of the free fixed point:
since the RG equations are perturbative, they are strictly valid only
within some sphere of small radius of the origin in $g$-space.  Within
certain domains of this space, the flows will be unstable, i.e. they
will exit the sphere of control, which indicates an instability
towards a new regime, and most likely an ordered state.  Within the
unstable regions of phase space, RG flows that begin very close to the
origin tend to converge toward particular unstable trajectories which
act as attractors, and are typically straight
``rays''\cite{PhysRevB.56.6569,balents1996weak}.  Below we follow
previous works that reformulate these rays to appear as fixed points,
by projecting the trajectories to a plane of constant value of one of
the parameters.  The stable ``fixed rays'' are expected to describe
the asymptotic limit of arbitrarily weak but non-zero bare
interaction, such that convergence to these rays is nearly perfect
before the sphere of control is exited.  One should keep in mind that
when the bare interactions are small but not arbitrarily so, the
deviations from these rays become important, and the physics will be
less universal and controlled more by the actual values of the
interactions, but the RG equations can still be employed.

Before considering the stable fixed rays, we note a few features of
these equations.  The $\beta$-functions for $g_1$ and $g_3$
contain an overall factor of $g_1$ and $ g_3$, respectively.  This
follows from symmetry: all terms except $g_1$ conserve spin at each
saddle point separately, and all terms except $g_3$ conserve the
number of electrons at each saddle point separately.  The conserving
interactions cannot generate a non-conserving one.  As a result, the
sign of $g_1$ and $g_3$ remain fixed through out the RG evolution.
One also notes that the $dg_2/dy>0$ and $dg_4/dy <0$ under the RG.
Thus an initially positive $g_2$ must remain positive, and an
initially negative $g_4$ remains negative.

To understand the physical meaning of the $g_i$, it is useful to
define interactions that parametrize particular channels of ordering,
e.g. they appear in the mean-field treatment below in
Sec.~\ref{sec:MFT}.  They were previously defined in Ref.~\cite{nandkishore2012chiral,LinYP2019}.
Here, we consider the interactions in the d-wave superconductivity
($G_{\rm dSC}=g_3-g_4$), s-wave superconductivity
($G_{\rm sSC}=-2 g_3-g_4$), real charge density
($G_{\rm rCDW}=-2 g_1+g_2-g_3$), orbital moment density
($G_{\rm iCDW}=-2 g_1+g_2+g_3$)~\footnote{Here, iCDW standards for ``imaginary charge density wave". However, note that the latter does not necessarily mean that the charge instability must break time reversal symmetry for a generic wave vector. But at wave vector $\veM=-\veM$ up to a reciprocal lattice vector, imaginary charge density must break time-reversal symmetry, and correspond to orbital moment density wave. For similar reasoning, we use iSDW for spin flux order.}, real spin density
($G_{\rm rSDW}=g_2+g_3$), spin flux order
($G_{\rm iSDW}=g_2-g_3$) channels. The interactions $G_a$ are defined
such that the $a$ instability develops only when $G_a>0$.

From the previous discussion, we can see a few features clearly.  Real
and imaginary parts of the CDW and SDW order parameters (OPs) are degenerate
if $g_3=0$.  This is because the umklapp interaction is the only one
transferring charge between saddle points, so that in its absence
there is a separate $U(1)$ charge rotation for each valley.
Similarly, the corresponding real and imaginary parts of the CDW and
SDW orders are degenerate when $g_1=0$.  This is because only $g_1$
violates separate spin conservation at each saddle, so that when $g_1=0$,
independent SU(2) rotations may be made for each flavor, which mixes
CDW and SDW orders.  Thus, we see that the sign of $g_3$ decides
between real and imaginary CDW/SDW, while the sign of $g_1$ decides
between CDW and SDW order.  


To proceed further, we determine the fixed rays and the pRG flow
trajectory near them.  We rewrite interactions as
$g_i = \gamma_i \mathsf{g}$, and choose $\mathsf{g}$ as one of the
interactions, which diverges as $\mathsf{g} \sim \frac{1}{y_c -y}$
along the fixed trajectory (as we verify afterwards). A proper
identification of $\mathsf{g}$ ensures that $\gamma_i$ tends to a
constant value $\gamma_i^*$ along the fixed trajectory, and
$\gamma_i^*= 0$ implies that the interaction $g_i$ either flows to
zero or diverges slower than $\frac{1}{y_c -y}$. We call it a fixed point hereafter.  The solutions to
$\gamma_i^*$ can be obtained by solving a set of algebraic equations
$\tilde\beta_i(\{\gamma\})=0$ for $i=1,2,3,4$, where
\begin{align}\label{eq:fixpRG}
\tilde\beta_i(\{\gamma\})=\dot{\gamma}_i=\frac{1}{\mathsf{g}} \left(\dot{g}_i-\gamma_i \dot{\mathsf{g}}\right).
\end{align}
Among those solutions, only the stable fixed point solutions are of physical interest, as they do not flow away under small perturbations. To examine the stability of a fixed point solution, we define a matrix $\mathsf{T}$ that is determined by the flow of $\tilde \beta_i$ at the fixed point, i.e.\ $\mathsf{T}_{ij}=\partial \tilde \beta_{i}/\partial \gamma_j|_{\{\gamma^*\}}$. The RG fixed point is stable only when all the eigenvalues of $\mathsf{T}$ are non-positive.

In addition to identifying the leading instabilities, the subleading
ones are also considered here for three reasons. \emph{First}, as
discussed above, when the interaction strengths are not truly
infinitesimal, flow to the unstable regime may occur before the fixed
ray is reached.  This may occur when bare couplings are small but not
too close to a fixed ray, and deviate from it in a particular
direction favoring a subleading instability.  \emph{Second}, the flow
may be also cut off by imperfect nesting and/or a non-zero chemical
potential (i.e. doping from the saddle point).  These effects limit
the convergence to the fixed ray, and at that point a different
instability might dominate.  \emph{Third}, a secondary instability may
develop at a lower temperature, even if the primary one occurs
first. To account for those possibilities, which depend upon the
microscopic details, we will list the leading two instabilities at the
RG fixed points.

To summarize, we list all the stable fixed point solutions for both repulsive and attractive bare interactions. In addition, we discuss an interesting semi-stable fixed point solution with only one weak unstable direction flowing out of the fixed point. 

When $g_2^{(0)}>0$, there are three (semi)stable fixed points, which
we take the liberty of denoting ``phases'' I, II, III -- this is an
abuse of terminology since these solutions really describe unstable
rays in the full phase space, which may not correspond to a unique phase.  As $g_2(y)$
must diverge as $\sim \frac{1}{y_c -y}$, we choose
$\mathsf{g}(y)=g_2(y)=\frac{1}{d_1(1+\gamma_3^2)(y_c-y)}$.
\begin{itemize}
\item[I.] When $g_3^{(0)} < 0$, $g_3$ flows to negative value at the stable RG fixed point, and we find $\gamma_1 = 0, \gamma_2 =1, \gamma_3 \approx -6.1, \gamma_4 \approx -5.5$. The subleading divergence of $g_1$ goes as $g_1(y) \sim (y_c-y)^{-\frac{2}{1+\gamma_3^2}}\sim \frac{1}{(y_c-y)^{0.05}}$. Note that as the flow of $g_1$ is subleading here, the sign of $g_1^{(0)}$ is not qualitatively important to determine the fixed trajectory. The leading instabilities are $G_{\rm sSC}=17.6\mathsf{g}, G_{\rm rCDW}=G_{\rm iSDW}=7.1\mathsf{g}$. 
\item[II.] When $g_1^{(0)}<0, g_3^{(0)}>0$, and for large enough $|g_1^{(0)}|$, the system may flow to a semi-stable fixed point, where there is only 1 weak unstable direction in the 4-dimensional parameter space defined by $\{g_1, g_2, g_3, g_4\}$. The fixed point solution reads $\gamma_1 \approx -46.7, \gamma_2 =1, \gamma_3 \approx 9.7, \gamma_4 \approx -4.4  $. The leading instabilities are $G_{\rm iCDW}=104.0\mathsf{g}, G_{\rm rCDW}=84.6\mathsf{g}$. 
\item[III.] When $g_1^{(0)}, g_3^{(0)} >0$, the stable RG fixed point has been discussed a lot in the literature~\cite{nandkishore2012chiral}. The fixed point solution gives $\gamma_1 = 0, \gamma_2 =1, \gamma_3 \approx 5.6, \gamma_4 \approx -10.0  $, where the subleading divergence of $g_1$ can be obtained as $g_1 (y) \sim (y_c-y)^{-\frac{2}{1+\gamma_3^2}} \approx \frac{1}{(y_c-y)^{0.06}}$. The leading divergent instabilities are $G_{\rm dSC}=15.6\mathsf{g}, G_{\rm rSDW}=G_{\rm iCDW}=6.6\mathsf{g}$.
\end{itemize}

In Fig.~\ref{fig:RG}, the phase diagram in the space of $g_1^{(0)} , g_3^{(0)}$ for $g_2^{(0)}>0$ is shown.

When $g_2^{(0)}<0$, $g_2$ may instead flow to zero. We find in this
way a fourth fixed ray, which we denote phase IV.  It is describe by letting
$\mathsf{g}=g_1(y)=-\frac{1}{2 d_1 (y_c-y)}$, and
$\gamma_1=1, \gamma_2=0, \gamma_3=0, \gamma_4=0$.   This solution
requires  $g_1^{(0)}<0, g_4^{(0)}>0$, and is only stable when
$g_3^{(0)}>0$.  Moreover, $g_3(y)$ is also
divergent and is only logarithmically smaller than $g_1$.   We find
$g_3(y)=\frac{1}{y_c-y}\left(\ln \frac{1}{y_c-y}\right)^{-1}$. The
leading instabilities are
$G_{\rm rCDW}=G_{\rm iCDW}=-2\mathsf{g}=\frac{1}{d_1(y_c-y)}$.   The
fixed point solution indicates that the rCDW and iCDW orders are
degenerate at the leading order, but they are split by a
logarithmically subdominant effect due to $g_3>0$ weakly in favor of
the iCDW, i.e. $G_{\rm iCDW}>G_{\rm rCDW}$.  

In passing, we note that purely electronic interactions generally give
repulsion for all couplings, i.e. $g_i^{(0)}>0$ with $i=1,2,3,4$. However, other
factors, such as orbital composition of the wave function near saddle
points, and electron-phonon coupling, may contribute to attraction for
certain $g_i^{(0)}$.   In App.~\ref{app:elph}, we consider the
effect of electron-phonon coupling, and show that the renormalization
to $g_i^{(0)}$ can be attractive for both $\delta g_1^{(0)}$ and
$\delta g_3^{(0)}$ or only $\delta g_1^{(0)}$, depending upon the
strength of the coupling and and phonon modes involved.

\begin{table}
\centering
\renewcommand{\arraystretch}{2}
\begin{tabularx}{0.48\textwidth}{|l|l|X|}
  \hline
  OP & Definition & Interaction strength $(G)$\\
  \hline
  rCDW &
  $N_{\alpha}=G_\textrm{rCDW}
  \frac{\abs{\epsilon_{\alpha\beta\gamma}}}{2\mathcal{N}}\sum_{\vec{q}}\expval{c^\dagger_{\beta
  \vec{q}}c_{\gamma \vec{q}}}$ &
  \multicolumn{1}{c|}{$-2g_1+g_2-g_3$}\\
  \hline
  iCDW & 
  $\phi_{\alpha}=G_\textrm{iCDW}
  \frac{\epsilon_{\alpha\beta\gamma}}{2i\mathcal{N}}\sum_{\vec{q}}\expval{c^\dagger_{\beta \vec{q}}c_{\gamma
  \vec{q}}}$ &
  \multicolumn{1}{c|}{$-2g_1+g_2+g_3$}\\
  \hline
  rSDW &
  $\vec{S}_{\alpha}=G_\textrm{rSDW}
  \frac{\abs{\epsilon_{\alpha\beta\gamma}}}{2\mathcal{N}}
  \sum_{\vec{q}}\expval{c^\dagger_{\beta \vec{q}}\frac{\vec{\sigma}}{2}c_{\gamma \vec{q}}}$ &
  \multicolumn{1}{c|}{$g_2+g_3$}\\
  \hline
  iSDW & 
  $\vec{\psi}_{\alpha}=G_\textrm{iSDW}
  \frac{\epsilon_{\alpha\beta\gamma}}{2i\mathcal{N}}
  \sum_{\vec{q}}\expval{c^\dagger_{\beta \vec{q}}\frac{\vec{\sigma}}{2}c_{\gamma \vec{q}}}$ &
  \multicolumn{1}{c|}{$g_2-g_3$}\\
  \hline
  sSC&
  $\Delta_s=G_\textrm{sSC}\frac{1}{\sqrt{3}\mathcal{N}}
  \sum_{\vec{q}}\expval{c_{\alpha\vec{q}  \downarrow}c_{\alpha
  -\vec{q}\uparrow}}$ & 
  \multicolumn{1}{c|}{$-2g_3-g_4$}\\
  \hline
  dSC &
  $\begin{aligned}
    \Delta_{xy}=&G_\textrm{dSC}\frac{1}{\mathcal{N}}\\
    &\times\sum_{\vec{q}}\expval{c_{\vec{q}\downarrow}D_{xy}c_{-\vec{q}\uparrow}}\\
    \Delta_{x^2-y^2}=&G_\textrm{dSC}\frac{1}{\mathcal{N}}\\
    &\times\sum_{\vec{q}}\expval{c_{\vec{q}\downarrow}D_{x^2-y^2}c_{-\vec{q}\uparrow}}\\
  \end{aligned}$ &
  \multicolumn{1}{c|}{$g_3-g_4$}\\
  \hline
\end{tabularx}
\caption{List of all bilinear order parameters (OPs) in the patch model. $\vec{\sigma}$ are the Pauli matrices in spin
space, and $\sum_{\vec{q}}\equiv\sum_{\abs{\vec{q}}<\Lambda}$. $D_{xy},D_{x^2-y^2}$ are matrices in the patch space
defined as $D_{xy}=\sqrt{1/2}\textrm{diag}\left(0,1,-1\right)$ and $D_{x^2-y^2}=\sqrt{2/3}\textrm{ diag}(1,-1/2,-1/2)$, and
$c_{\vec{q}\sigma}=(c_{1\vec{q}\sigma},c_{2\vec{q}\sigma},c_{3\vec{q}\sigma})$.}
\label{eq:OPtable}
\end{table}

\section{Mean-field theory}\label{sec:MFT}
As shown in Sec.~\ref{sec:RG}, we found instabilities to
superconducting, charge density wave (rCDW), orbital moment (iCDW) and
spin density wave (SDW) states.  In all three AV$_3$Sb$_5$ materials,
experiments have observed charge density wave order (rCDW order)
setting as the first instability of the symmetric state at a
$T_c \sim 90K$, and superconductivity only at much lower temperatures.
Hence, in this section, we neglect superconductivity, and discuss the
ways that the remaining instabilities may lead to the specific rCDW
order observed experimentally in the AV$_3$Sb$_5$ materials.  Notably,
we find that rCDW order can be induced even if rCDW is not the primary order
parameter.  Hence we study the formation of rCDWs both when the rCDW
is and is not the primary instability, and discuss the differences in
the resulting properties.  We carry out the study using mean field
theory.  

The renormalization group analysis tells us that in phase I the rCDW
is a leading instability.
In Sec.~\ref{sec:RCDWMFT}., we will study this case by calculating a
mean-field theory in the rCDW channel. In Sec.~\ref{sec:ICDWRCDWMFT},
we consider a iCDW-rCDW coupled mean field theory, which is relevant
to phases II and IV.  Here we discuss how rCDW can be induced when the
iCDW is the primary order parameter. Finally, in
Sec.~\ref{sec:RSDWRCDWMFT}, we consider the situation with a leading
rSDW instability, relevant to phase III, and show how it may induce
rCDW order.  

\subsection{Complex CDW free energy}
\label{sec:complex-cdw-free}

We  first obtain the Landau free
energy including both  rCDW and iCDW
order parameters, defined as $N_\alpha$, $\phi_\alpha$ with $\alpha=1,2,3$ labeling the interpatch momentum transfer
$\ve{Q}_\alpha$ (see Fig.~\ref{fig:latticeBZ} (b)), respectively. 
The patch-model interaction given in Eq. \eqref{eq:interaction} can be rewritten in the form of a rCDW interaction and an iCDW interaction:
\begin{align}
  H_\textrm{rCDW}&=-\frac{\mathcal{N} G_\textrm{rCDW}}{2}\sum_{\alpha}\hat{\rho}_{\textrm{rC},\alpha}\hat{\rho}_{\textrm{rC},\alpha},\non\\
  H_\textrm{iCDW}&=-\frac{\mathcal{N} G_\textrm{iCDW}}{2}\sum_{\alpha}\hat{\rho}_{\textrm{iC},\alpha}\hat{\rho}_{\textrm{iC},\alpha},
\end{align}
where $G_\textrm{r/iCDW}$ is the r/iCDW interaction strength defined in Sec.~\ref{sec:RG} and
$\hat{\rho}_{\textrm{rC},\alpha}=\frac{\abs{\epsilon_{\alpha\beta\gamma}}}{2\mathcal{N}}
\sum_{\abs{\vec{q}}<\Lambda}c^\dagger_{\beta \vec{q}}c_{\gamma \vec{q}}$, $\hat{\rho}_{\textrm{iC},\alpha}=\frac{\epsilon_{\alpha\beta\gamma}}{2\iu\mathcal{N}}
\sum_{\abs{\vec{q}}<\Lambda}c^\dagger_{\beta \vec{q}}c_{\gamma \vec{q}}$ is
the rCDW and iCDW operators with momentum $\bm{Q}_\alpha=\bm{M}_\beta-\bm{M}_\gamma$.  Using the Hubbard-Stratonovich transformation we decouple the
interactions in the two channels and then integrate out the Fermionic degrees of freedom. This gives us the free
energy as a function of the rCDW, iCDW order parameters,
$N_1,N_2,N_3,\phi_1,\phi_2,\phi_3$:
\begin{equation}
  F_\textrm{CDW}=\frac{\mathcal{N}}{2G_\textrm{rCDW}}\sum_{\alpha=1}^3N_\alpha^2+\frac{\mathcal{N}}{2G_\textrm{iCDW}}\sum_{\alpha=1}^3\phi_\alpha^2-(\Tr\log\mathcal{G}^{-1}),
  \label{eq:irCDWFreeEnergy}
\end{equation}
where $\mathcal{G}^{-1}$ is defined as
\begin{align}
  &\mathcal{G}^{-1}(i\omega_n,\vec{q};\{N_\alpha,\phi_\alpha\})\notag\\
  &=
  \begin{bmatrix}
    -i\omega_n+\varepsilon_1(\vec{q})&-(N_3-\iu \phi_3)/2&-(N_2+\iu \phi_2)/2\\
    -(N_3+\iu \phi_3)/2 & -i\omega_n+\varepsilon_2(\vec{q})&-(N_1-\iu \phi_1)/2\\
    -(N_2-\iu \phi_2)/2 & -(N_1+\iu \phi_1)/2 & -i\omega_n+\varepsilon_3(\vec{q})
  \end{bmatrix}.
  \label{eq:inverseirCDWGF}
\end{align}
Expanding the order parameter fields in Eq.~\eqref{eq:inverseirCDWGF}  perturbatively, the free energy in terms of the complex charge density order parameter $\Delta_\alpha=N_\alpha +\iu \phi_\alpha=\abs{\Delta_\alpha}\eu^{\iu \theta_\alpha}$ is
\begin{align}
f_{\rm CDW}=&\frac{r_N+r_\phi}{2}\sum_\alpha |\Delta_\alpha|^2+\frac{r_N-r_\phi}{2}\sum_\alpha \abs{\Delta_\alpha}^2 \cos 2\theta_\alpha\non\\
&+K_2 \abs{\Delta_1}\abs{\Delta_2}\abs{\Delta_3}\cos(\theta_1+\theta_2+\theta_3)\non\\
&+K_4 \left(\sum_\alpha \abs{\Delta_\alpha}^2\right)^2+(K_3-2K_4)\sum_{\alpha<\beta} \abs{\Delta_\alpha}^2\abs{\Delta_\beta}^2\non\\
&+\mathcal{O}(\Delta^5).
\label{eq:riCDWexpansion}
\end{align}
Here, $r_N=\frac{1}{2G_\textrm{rCDW}}+K_1, r_\phi=\frac{1}{2G_\textrm{iCDW}}+K_1$, and $K_1,\cdots K_4$ are functions of temperature and chemical potential, that we will discuss in detail in Sec.~\ref{sec:RCDWMFT}. For simplicity, we first consider the case $K_3-2K_4<0$, which favors 3Q CDW. The configuration of $\theta_\alpha$ that minimizes the free energy depends on the sign of $r_N-r_\phi$:
\begin{itemize}
\item When $r_N<r_\phi$, i.e. $G_{\rm rCDW}>G_{\rm iCDW}>0$, minimizing the 2nd term requires $\theta_\alpha= n_\alpha\pi$, with $n_\alpha\in \mathbb{Z}$. By choosing the proper $n_\alpha$, the minimization of cubic term can be readily achieved. For any $\theta_\alpha= n_\alpha\pi$, the iCDW order parameter must vanish, so the ground state only contains rCDW order. 
\item When $r_N=r_\phi$, i.e. $G_{\rm rCDW}=G_{\rm iCDW}$, the second term vanishes, and the minimization of the cubic term gives a continuously degenerate ground state manifold. To lift the degeneracy, other perturbations should be considered.
\item When $r_N>r_\phi$, i.e. $0<G_{\rm rCDW}<G_{\rm iCDW}$, the second quadratic and cubic terms cannot be minimized simultaneously. As a result, $\theta_\alpha \neq n_\alpha \pi/2$, with $n_\alpha\in \mathbb{Z}$. This indicates that both $N_\alpha$ and $\phi_\alpha$ are non-zero. To analyze the Free energy with iCDW as the leading instability, the iCDW-rCDW coupling must be considered. 
\end{itemize}

Below, we discuss the above three scenarios and then the rSDW-rCDW coupling scenario.

\subsection{rCDW Mean-field theory}\label{sec:RCDWMFT}
At the RG fixed point corresponding to phase I ($g_3^{(0)}<0$), the rCDW is the leading density wave instability. As argued above, it is enough to consider only the rCDW order parameters. The mean field free energy becomes
\begin{equation}
  F_\textrm{rCDW}=\frac{\mathcal{N}}{2G_\textrm{rCDW}}\sum_{\alpha=1}^3N_\alpha^2-(\Tr\log\mathcal{G}^{-1}),
  \label{eq:fullCDWFreeEnergy}
\end{equation}
where $\mathcal{G}^{-1}$ is defined as
\begin{align}
  &\mathcal{G}^{-1}(i\omega_n,\vec{q};\{N_\alpha\})\notag\\
  &=
  \begin{bmatrix}
    -i\omega_n+\varepsilon_1(\vec{q})&-N_3/2&-N_2/2\\
    -N_3/2 & -i\omega_n+\varepsilon_2(\vec{q})&-N_1/2\\
    -N_2/2 & -N_1/2 & -i\omega_n+\varepsilon_3(\vec{q})
  \end{bmatrix}.
  \label{eq:inverseCDWGF}
\end{align}
As a function of $N_\alpha$, the rCDW free energy has the symmetry of the tetrahedral point group, $T_d$. This can be
deduced by applying the symmetry operations of the full space group of the lattice, $P6/mmm$, to the definitions of the
rCDW order parameter. Using our knowledge of this symmetry, we can determine what solutions of the free energy are
possible. In the cases that are physically relevant, the solutions of the mean-field theory belong in either the 3Q$+$,
3Q$-$, or 1Q rCDW configurations which are classes of rCDW states with directions:
\smallskip
\begin{center}
\begin{tabular}{l l}
  3Q$+$: & $\{(111),(1\bar{1}\bar{1}),(\bar{1}1\bar{1}),(\bar{1}\bar{1}1)\}$,\\
  3Q$-$: & $\{(\bar{1}\bar{1}\bar{1}),(\bar{1}11),(1\bar{1}1),(11\bar{1})\}$,\\
  1Q: & $\{(100),(\bar{1}00),(010),(0\bar{1}0),(001),(00\bar{1})\}$.
\end{tabular}
\end{center}
\smallskip
Indeed, by numerically solving the full free energy we will see that the solutions belong in either the 3Q$\pm$ or 1Q
rCDW classes.  A detailed discussion is provided in App.~\ref{app:morseDiscussion}. The rCDW patterns of the
3Q$\pm$ and 1Q states are shown in Fig.~\ref{fig:rCDWPattern}.

By assuming we are near the rCDW transition where the rCDW order parameters are sufficiently small, we can expand the
free energy to fourth order in $N_\alpha$. The resulting Landau theory is
\begin{align}
  &f_\textrm{rCDW}=\left(\frac{1}{2G_\textrm{rCDW}}+K_1\right)\sum_{\alpha}N_\alpha^2+K_2N_1N_2N_3\notag\\
  &\quad+K_4\left(\sum_{\alpha}N_\alpha^2\right)^2+(K_3-2K_4)\sum_{\alpha<\beta}N_\alpha^2N_\beta^2+\mathcal{O}(N^5),
  \label{eq:rCDWLandauTheory}
\end{align}
where $f_\textrm{rCDW}$ is the free energy density. The definitions of the coefficients $K_1,\cdots,K_4$ which are
functions of chemical potential and temperature are given in App.~\ref{app:landaucoefficients}. The coefficients can be
evaluated asymptotically in the limit $\mu,k_BT\ll t\Lambda^2$ and can also be found in App.
\ref{app:landaucoefficients}.

\begin{figure}[htp]
  \begin{center}
    \includegraphics[width=\linewidth]{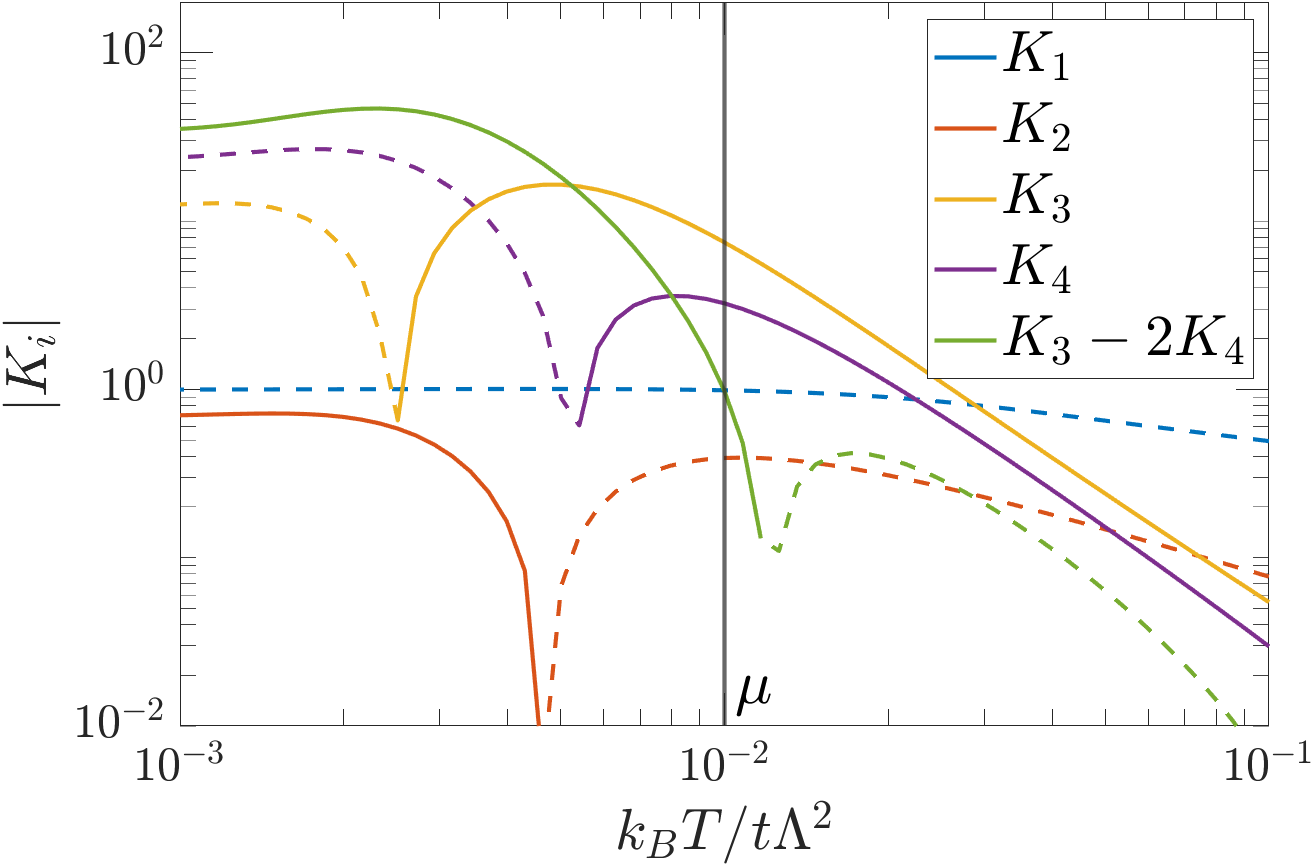}
  \end{center}
  \caption{Absolute value of the Landau theory coefficients, $K_1,\cdots,K_4$ evaluated at $\mu/t\Lambda^2=10^{-2}$ for the
  case of perfect nesting ($a=9t/4\,, b= 3t/4$). Solid lines indicate positive values and dashed lines indicate negative
  values.  $K_2,K_3,K_4$ change sign at $\mu/k_BT\sim2.14,4.05,1.91$ respectively.}
  \label{fig:LandauCoefficients}
\end{figure}

Fig.~\ref{fig:LandauCoefficients} shows the temperature dependence of these coefficients when $\mu/t\Lambda^2=0.01$.
Notice that the coefficients exhibit a change in behavior at the cross-over temperature $k_B T\sim\mu$.  In addition, the $K_2,K_3,K_4$
coefficients change sign near the cross-over temperature. When the quartic coefficients become negative, the
Landau theory expression given by Eq.  \eqref{eq:rCDWLandauTheory} becomes unstable, but in those regions stability
can be restored by including sixth-order terms to the free energy.

Eq. \eqref{eq:rCDWLandauTheory} has a third order term, $K_2$, that couples all three rCDW order parameters.  This is
allowed by symmetry since this term is even under time-reversal symmetry and the sum of the three nesting wave vectors
satisfies $\sum_\alpha \ve{Q}_\alpha\equiv0$. This term introduces a preference for 3Q$+$ or 3Q$-$ rCDW states depending on
the sign of $K_2$. On the other hand, when $K_3-2K_4>0$, the fourth order term sets a preference for 1Q rCDW states.
When $K_3-2K_4<0$, this fourth order term prefers the 3Q$\pm$ states equally.

When $a/b=3$, Fig.~\ref{fig:LandauCoefficients} shows that $K_2<0$ when $\mu/k_BT\lesssim2.14$ in the asymptotic limit.
In addition, $K_3-2K_4<0$ for $\mu\ll k_BT$. Hence, in this region, all terms in the rCDW Landau theory prefer the 3Q$+$
rCDW state. The transition to the 3Q$+$ rCDW state must be a first-order transition because the free energy can become
negative before the second-order term vanishes due to the third-order term.  One thing to note is that the rCDW order
parameter is not necessarily small near the first-order rCDW transition, so the results of the Landau theory must be
treated with caution. We can avoid this issue by numerically solving the full free energy.

The full free energy defined in Eqs.  \eqref{eq:fullCDWFreeEnergy}, \eqref{eq:inverseCDWGF} has four tunable parameters:
temperature $T$, chemical potential $\mu$, the rCDW interaction strength $G_\textrm{rCDW}$, and the nesting ratio $a/b$.
As explained in Sec.~\ref{sec:Model}, we only need to consider the case $a,b>0$. Given this condition, we can
introduce a convenient reparametrization,
\begin{align}
  a=&\frac{\delta+\sqrt{3+\delta^2}}{\sqrt{3}}t,\notag\\
  b=&\frac{-\delta+\sqrt{3+\delta^2}}{\sqrt{3}}t,
  \label{eq:tdeltaparametrization}
\end{align}
where $t>0\textrm{ and }\delta\in\mathbb{R}$. We see that $t$ represents the bandwidth of the saddle point band and
$\delta$ represents the degree of nesting.  Under this parametrization, $a/b\geq1$ for $\delta\geq0$ and $a/b<1$ for
$\delta<0$. In particular, $a/b=3$ (perfect nesting) for $\delta=1$.

\begin{figure*}[t]
  \centering
  \subfigure[]{\includegraphics[width=0.30\linewidth]{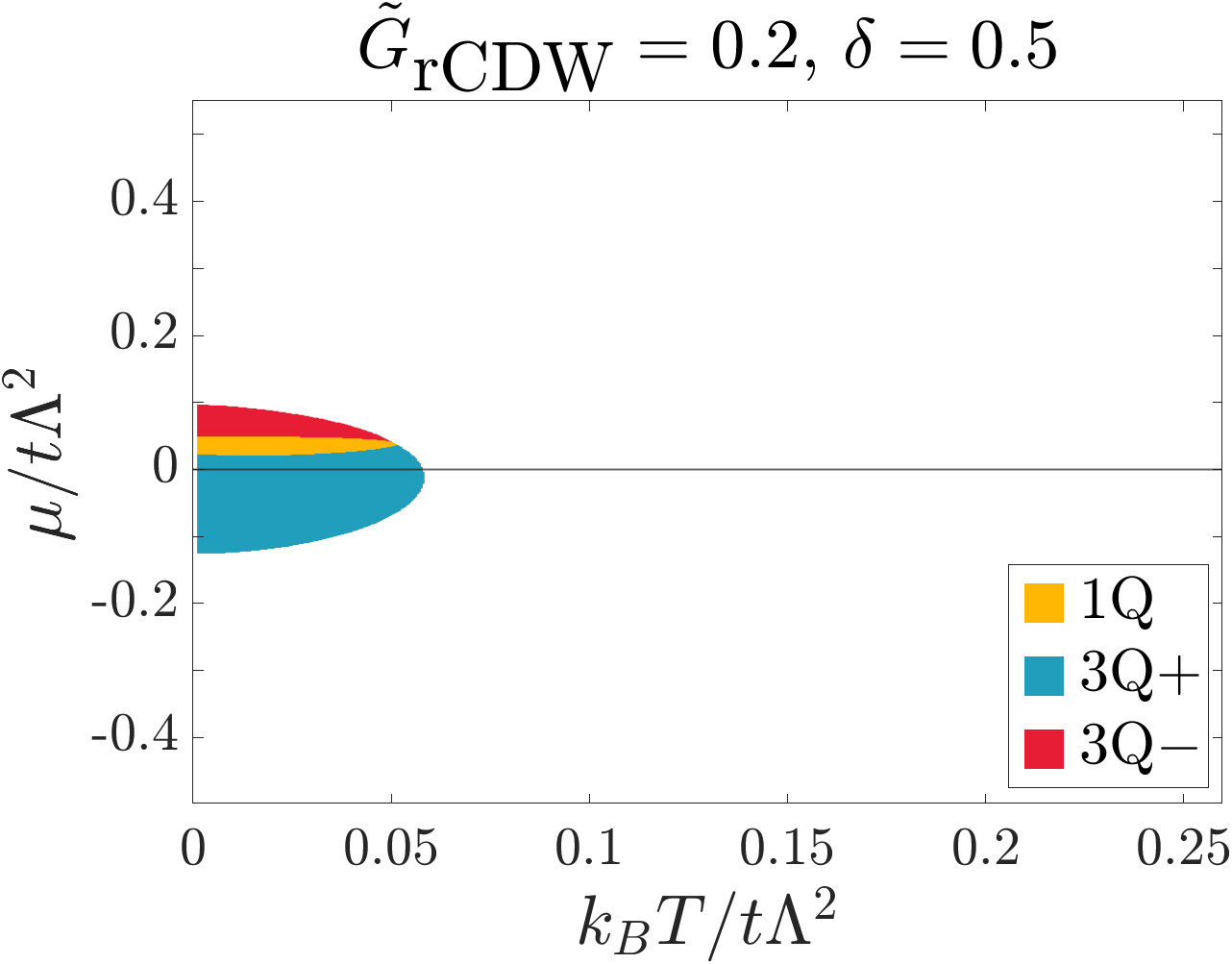}\label{fig:phasediagramsa}}
  \subfigure[]{\includegraphics[width=0.30\linewidth]{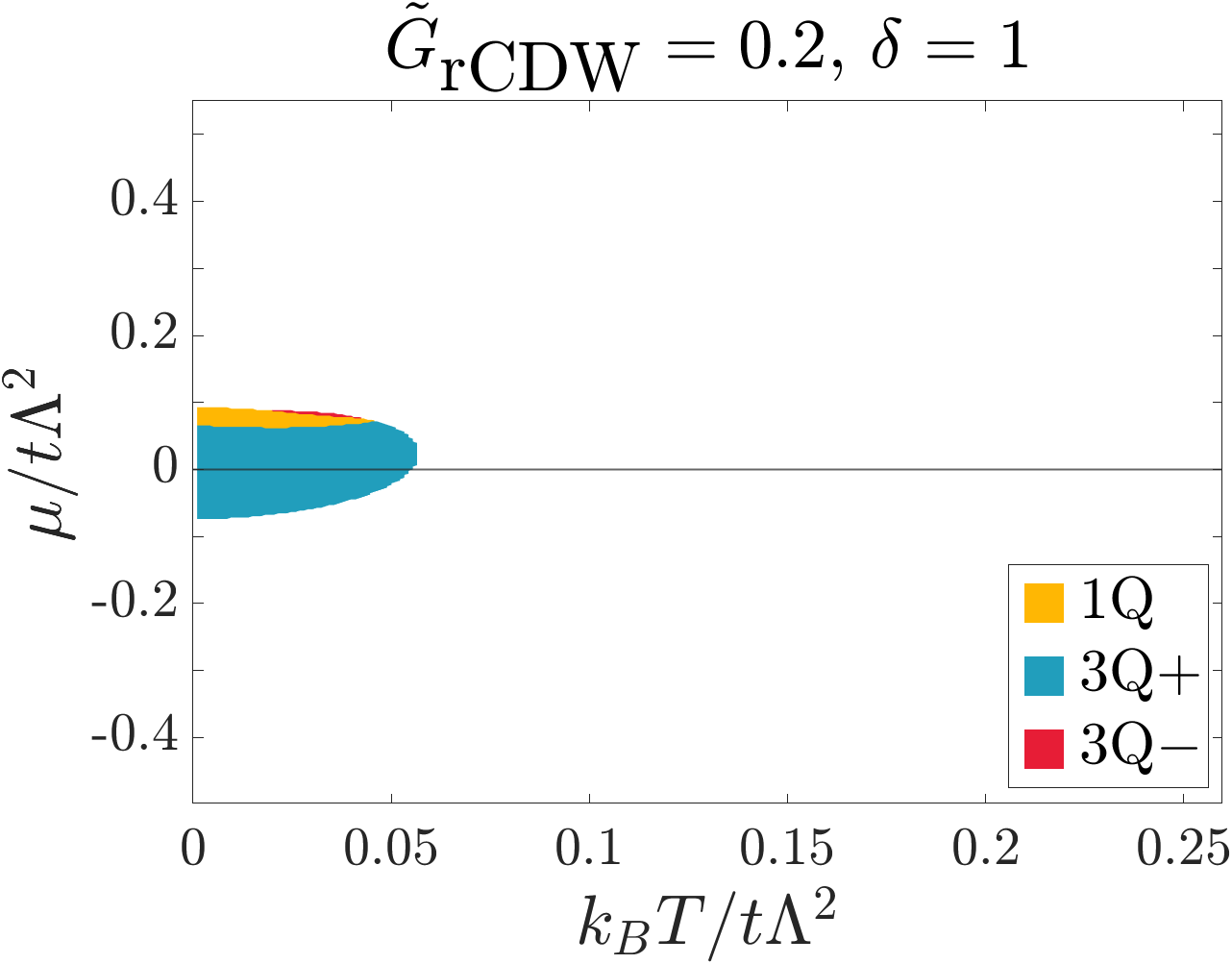}\label{fig:phasediagramsb}}
  \subfigure[]{\includegraphics[width=0.30\linewidth]{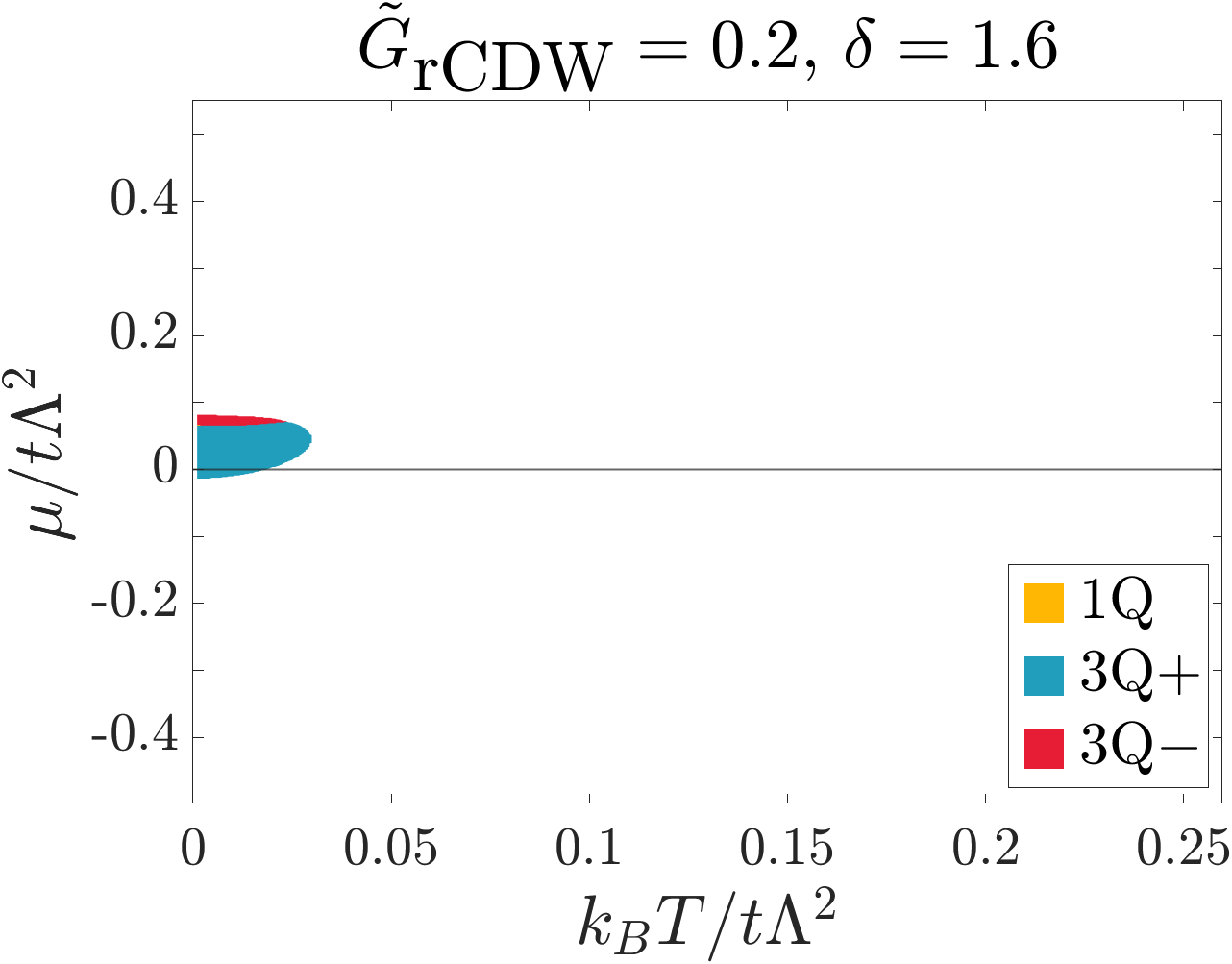}\label{fig:phasediagramsc}}\\
  \subfigure[]{\includegraphics[width=0.30\linewidth]{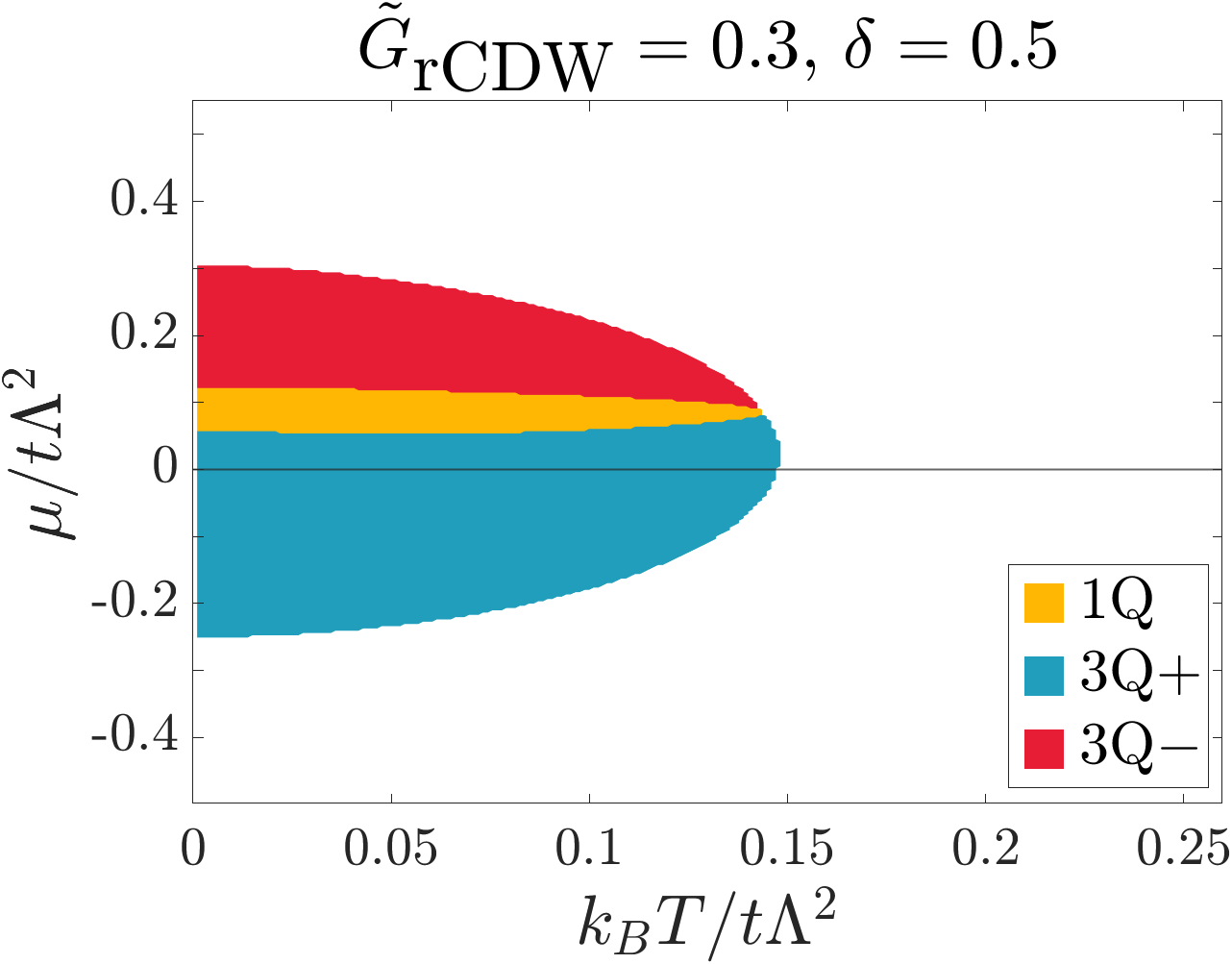}\label{fig:phasediagramsd}}
  \subfigure[]{\includegraphics[width=0.30\linewidth]{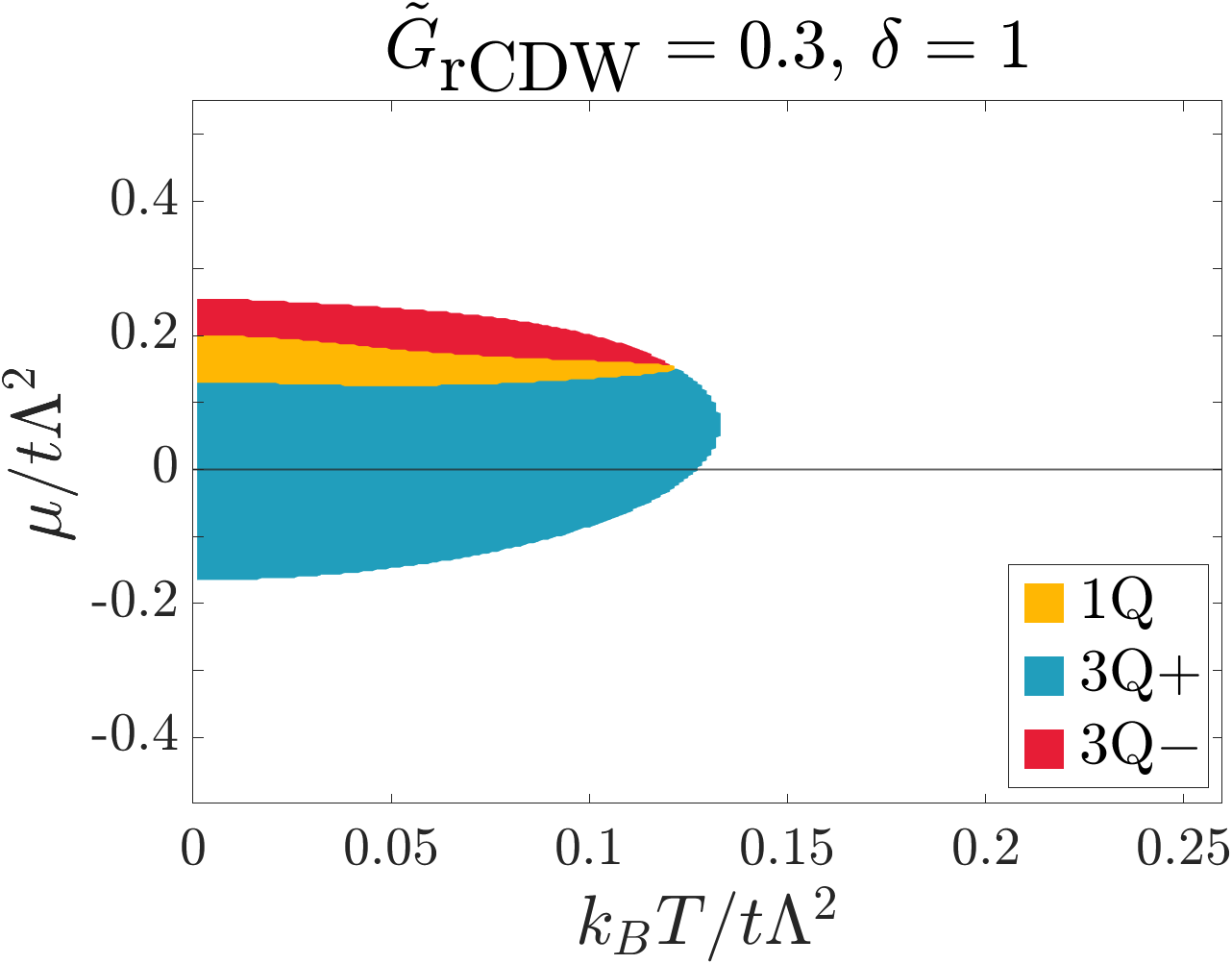}\label{fig:phasediagramse}}
  \subfigure[]{\includegraphics[width=0.30\linewidth]{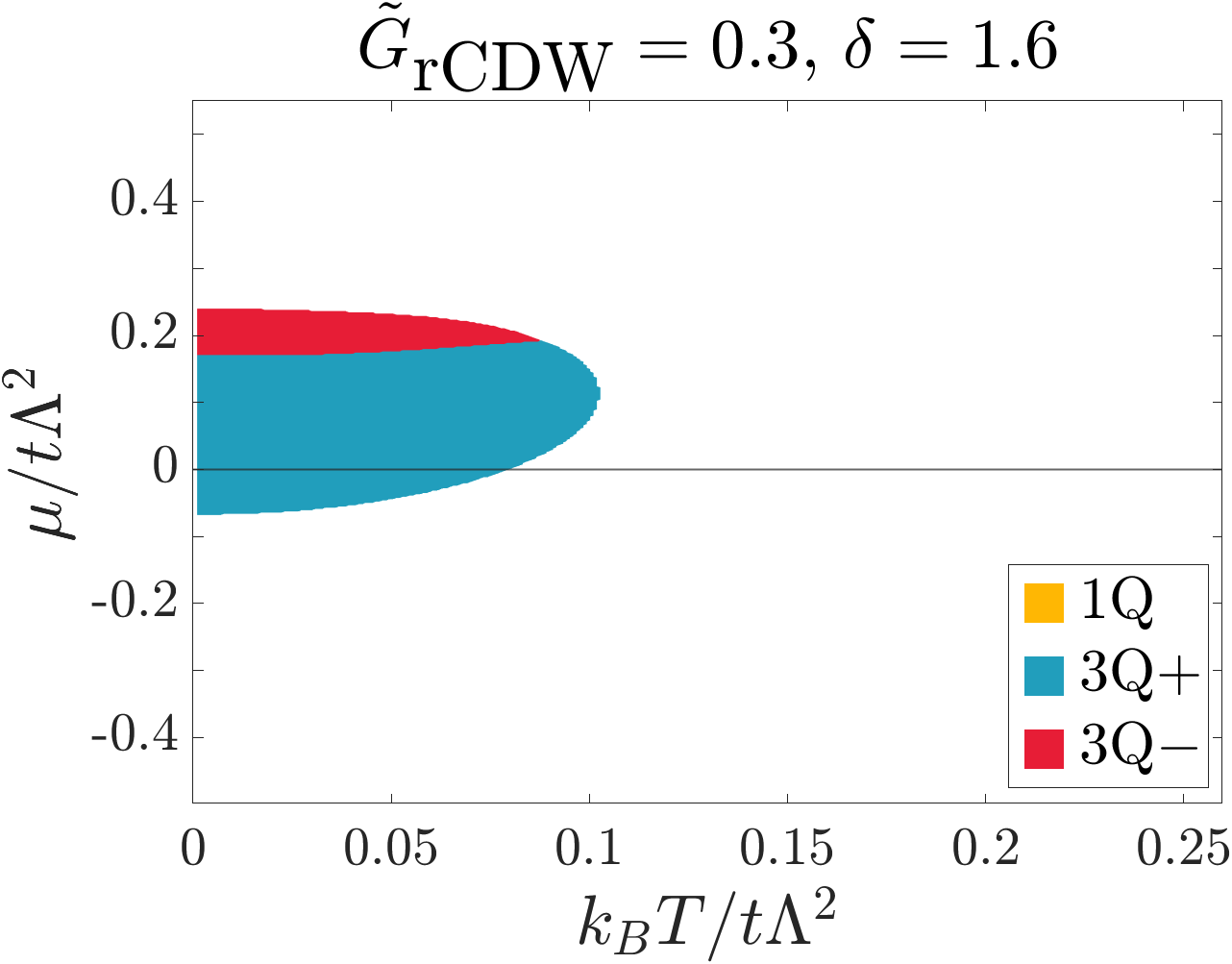}\label{fig:phasediagramsf}}\\
  \subfigure[]{\includegraphics[width=0.30\linewidth]{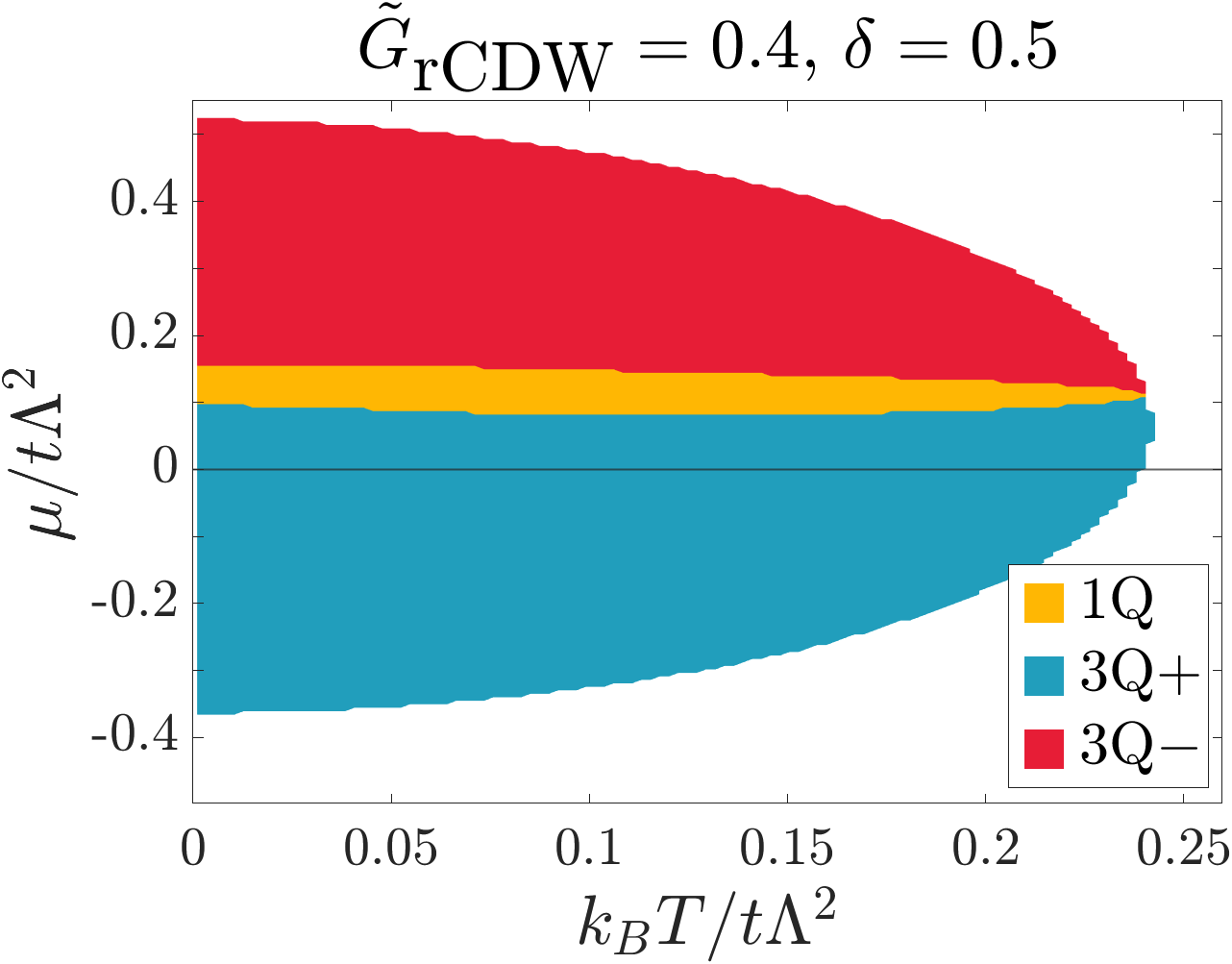}\label{fig:phasediagramsg}}
  \subfigure[]{\includegraphics[width=0.30\linewidth]{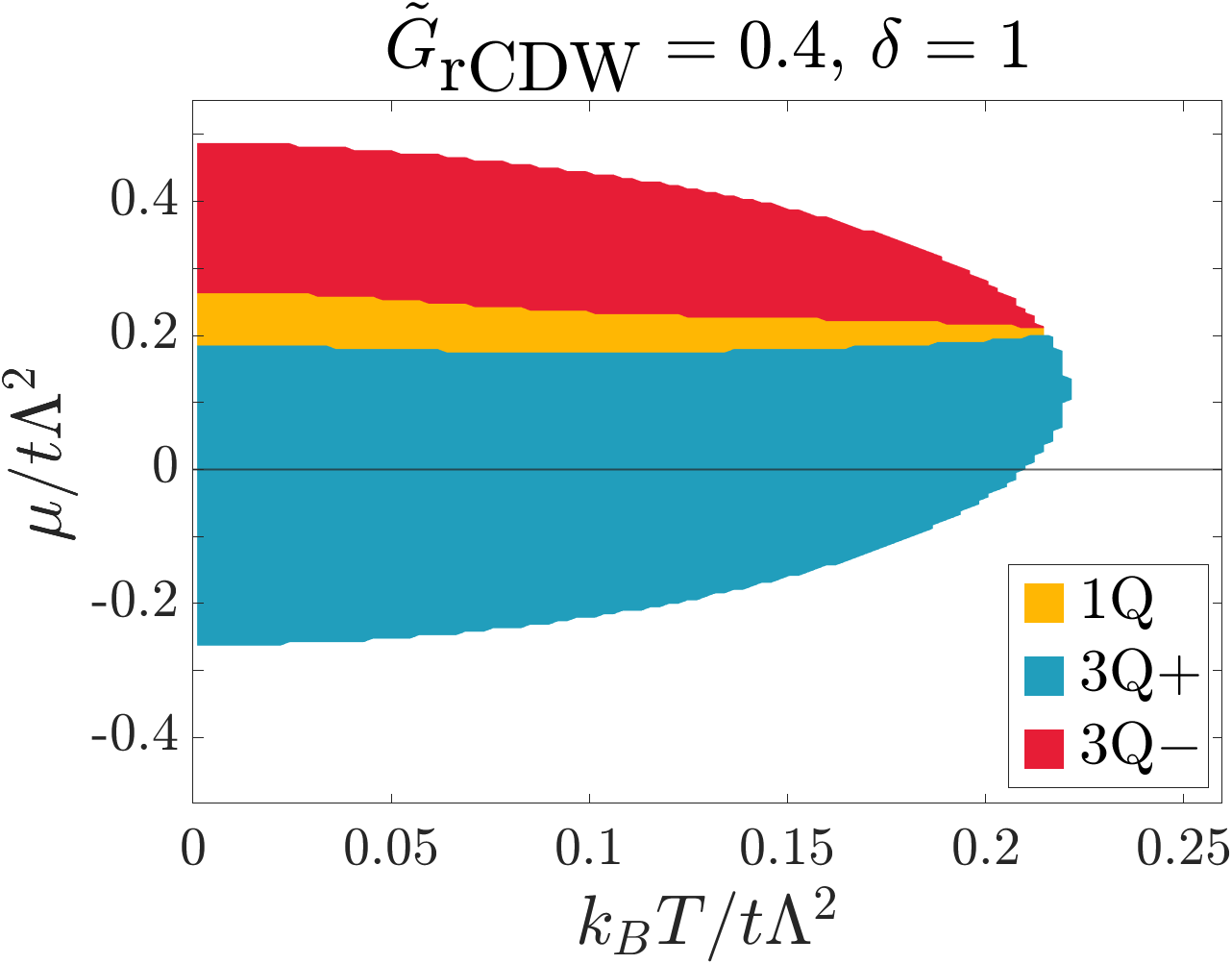}\label{fig:phasediagramsh}}
  \subfigure[]{\includegraphics[width=0.30\linewidth]{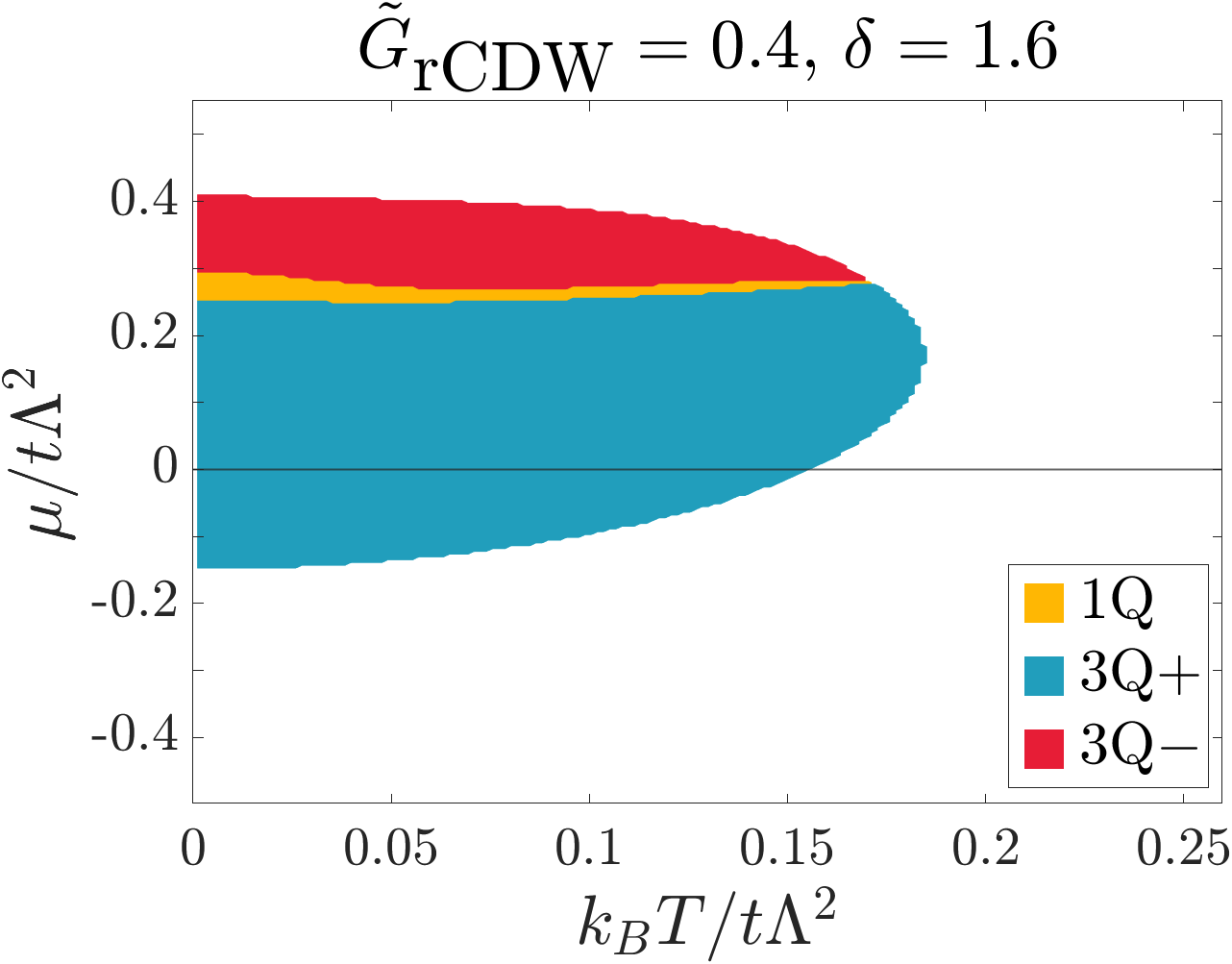}\label{fig:phasediagramsi}}
  \caption{rCDW phase diagrams calculated in the $k_BT$--$\mu$ plane.
  $\tilde{G}_\textrm{rCDW}=G_\textrm{rCDW}/t\Lambda^2$ is defined as the dimensionless rCDW strength. $\delta$
  parametrizes nesting as defined in Eq.  \eqref{eq:tdeltaparametrization}.  The phase diagrams correspond to the case
  $t,\delta>0$ ($a,b>0\textrm{ and }a/b>1$). The phase diagrams for $t,\delta<0$ ($a,b>0\textrm{ and }a/b<1$) are
  related to these phase diagrams under the exchange $\textrm{3Q}+\leftrightarrow\textrm{3Q}-$ and reflection in the
  $\mu=0$ axis.  }
  \label{fig:phasediagrams}
\end{figure*}

For different values of $G_\textrm{rCDW}$ and $\delta$, we generate a phase diagram in the $T-\mu$ plane. The results
shown in Fig.~\ref{fig:phasediagrams} are representative samples of the the entire parameter space where the continuum
model holds---$k_BT\,,\mu\,,G_\textrm{rCDW}\ll t\Lambda^2$. We see that when the system is doped to the saddle point
($\mu=0$) or below it, only the 3Q$+$ rCDW forms. As you dope the system above the saddle point, 1Q and/or 3Q$-$ rCDW
regions can emerge. In addition, for $\delta=1.6$, the 1Q rCDW region is small and vanishes for sufficiently small
$G_\textrm{rCDW}$ as seen in Figs.~\ref{fig:phasediagramsc},~\ref{fig:phasediagramsf}.  Similarly, at perfect nesting and more generally
$\delta\sim1$, the 3Q$-$ rCDW region vanishes for sufficiently small $G_\textrm{rCDW}$ as seen in Fig.
\ref{fig:phasediagramsb}.  Clearly, the 3Q$+$ rCDW region is largest in all phase diagrams which is consistent with
the Landau theory discussed above.

\subsection{iCDW-rCDW mean field theory}
\label{sec:ICDWRCDWMFT}
Now, we consider how a fundamental iCDW order can induce subsidiary
rCDW order. To do so, we consider a iCDW-rCDW coupled mean field
theory.  We are particularly interested in the regime, motivated by
the RG results for phases II and IV, in which iCDW
and rCDW are close in energy and can compete.


It is convenient to express Eq.~\eqref{eq:riCDWexpansion} in terms $N_\alpha$ and $\phi_\alpha$:
\begin{align}
 &f_{\rm CDW}=r_\phi\sum_{\alpha=1}^3\phi_\alpha^2+r_N\sum_{\alpha=1}^3N_\alpha^2\notag\\
  &+K_2\left(N_1 N_2 N_3 -\phi_2 \phi_3 N_1 - \phi_1 \phi_3 N_2- \phi_2 \phi_1 N_3\right)\notag\\
  &+K_4 \left(\sum_{\alpha=1}^3\phi_\alpha^2+N_\alpha^2\right)^2\non\\
 &+(K_3-2K_4)\sum_{\alpha<\beta}(\phi_\alpha^2+N_\alpha^2)(\phi_\beta^2+N_\beta^2)+\mathcal{O}(\phi^6, N^5).
  \label{eq:iCDWLandauTheory}
\end{align}
Recall that $r_\phi=\left(\frac{1}{2G_\textrm{iCDW}}+K_1\right)$,
$r_N=\left(\frac{1}{2G_\textrm{rCDW}}+K_1\right)$.  We assume $G_{\rm
  iCDW} > G_{\rm rCDW}>0$, so that iCDW is the leading
instability. Note that time-reversal symmetry forbids the term cubic
in $\phi$, i.e.\ $\phi_1 \phi_2 \phi_3$. As a result, the order
parameter manifold for $\phi_\alpha$ alone has cubic symmetry
$O_h$. In addition, the 3Q$\pm$ classes which were distinct
for the case of rCDW are now related under time-reversal symmetry and part of the larger 3Q class which is defined as the union
of the 3Q$\pm$ classes.
Close to and below the transition temperature for iCDW, the iCDW-rCDW coupling can be treated as a perturbation.

First, we consider the iCDW Landau theory. The order parameter
manifold for $\phi_\alpha$ is determined by quartic terms proportional
to $K_3$ and $K_4$. The first term $K_4 (\sum_{\alpha=1}^{3}
\phi^2_\alpha)^2$ is positive definite and isotropic, so the selection
of the ground state configuration is determined by the term
$(K_3-2K_4)\sum_{\alpha<\beta}\phi_\alpha^2\phi_\beta^2$. Specifically,
(i) when $K_3> 2K_4$, the 1Q iCDW state is favored, (ii) when $K_3 =
2K_4$, the $\phi_\alpha$ ground state configuration is degenerate at
quartic order, and requires higher order terms to break the
degeneracy, (iii) when $K_3< 2K_4$, the 3Q iCDW is favored. From
the asymptotic solution shown in Fig.~\ref{fig:LandauCoefficients} at $\mu/t\Lambda^2=0.01$, we
see that when $k_B T_c/t\Lambda^2\lesssim 0.01$, and the 1Q iCDW is stable. When $k_B T/t\Lambda^2\gtrsim 0.01$, $K_3-2K_4<0$. Thus 3Q iCDW develops if $T_c$ sits in this range of temperature, but it may become unstable to 1Q iCDW as temperature lowers. On the other hand, when $\mu\ll k_B T$, we find $K_3-2K_4<0$ in the low temperature regime when $k_B T/t \Lambda^2 < 0.025$ (see App.~\ref{app:landaucoefficients}). Thus 3Q iCDW should be stable at $\mu \sim 0$. 
This transition is a continuous phase transition in contrast to the rCDW case because of the absence of a third-order term.
The real space pattern for 1Q and 3Q iCDW orders are presented in Sec.~\ref{sec:icdwpattern}.

Next, we substitute the iCDW  order parameter in Eq.~\eqref{eq:iCDWLandauTheory} with the iCDW saddle point solution,
$|\phi_\alpha|=\phi^*$, and obtain
the free energy for rCDW OPs. For 1Q iCDW, no rCDW can be induced. For 3Q iCDW, as the cubic term $K_2$
contains term linear in $N_\alpha$ through coupling to two iCDW OPs at another two momenta, considering up to quadratic term
in $N$, we find $\{N_1, N_2, N_3\}=N_0 \{1, \pm 1 , \pm 1\}$ up to any
permutations of the signs between $N_\alpha$, where
$N_0\sim\frac{K_2 \phi^{*2}}{(T-T_{\rm rCDW})}$. Here,
$T-T_{\rm rCDW} \sim \left(\frac{1}{2 G_{\rm rCDW}}+K_1 \right)>0$. So
the sign of $N_0$ is determined by the sign of $K_2$. Since $K_2<0$
for $\mu/T\lesssim 2.14$, $N_0<0$, the induced rCDW order must be the
3Q$-$ one. As temperature further lowers to $T<T_{\rm rCDW}$, the
quartic term in $N_\alpha$ should be included so that the free energy is stable. We checked that the rCDW order remains the 3Q$-$ one for
$K_2<0$ and vice versa.

\subsection{rSDW-rCDW Mean-field theory}\label{sec:RSDWRCDWMFT}
In phase III of the RG phase diagram, the rSDW is the leading density
wave instability.  Here, we consider how rCDW order can emerge as a
subsidiary order through rSDW-rCDW coupling.  By rewriting the full
interaction, Eq. \eqref{eq:interaction}, using the rSDW operators the rSDW interaction term is
\begin{equation}
  H_\textrm{rSDW}=-\frac{\mathcal{N}
  G_\textrm{rSDW}}{2}\sum_\alpha\hat{\vec{\rho}}_{\textrm{rS},\alpha}\cdot\hat{\vec{\rho}}_{\textrm{rS},\alpha},
\end{equation}
where $\hat{\vec{\rho}}_{\textrm{rS},\alpha}=\frac{\abs{\epsilon_{\alpha\beta\gamma}}}{2\mathcal{N}}
\sum_{\abs{\vec{q}}<\Lambda}c^\dagger_{\beta\vec{q}}\frac{\vec{\sigma}}{2}c_{\gamma \vec{q}}$
is the spin density operator. Assuming the interaction in rCDW channel is also attractive, but much weaker than that of
rSDW, we can decouple the interactions in the rSDW and rCDW channel using a Hubbard-Stratonovich transformation,
integrate out the Fermionic degrees of freedom, and expand the resulting mean field free energy to fourth order in the
rSDW and rCDW order parameters.
This gives us the rSDW free energy (which may be added to the rCDW free energy in Eq. \eqref{eq:rCDWLandauTheory}):
\begin{align}
  f_\textrm{rSDW}=&\left(\frac{1}{2G_\textrm{rSDW}}+\frac{K_1}{4}\right)\sum_\alpha\abs{\vec{S}_\alpha}^2\notag\\
&+\frac{K_2}{4}\left(N_1\vec{S}_2\cdot\vec{S}_3+N_2\vec{S}_3\cdot\vec{S}_1+N_3\vec{S}_1\cdot\vec{S}_2\right)\notag\\
&+\frac{K_3}{16}\sum_{\alpha<\beta}\abs{\vec{S}_\alpha}^2\abs{\vec{S}_\beta}^2+\frac{K_4}{16}\sum_\alpha\abs{\vec{S}_\alpha}^4+\notag\\
&+K_5\sum_{\alpha<\beta}\abs{\vec{S}_\alpha\cdot\vec{S}_\beta}^2+K_6\left(\vec{S}_1\cdot\vec{S}_2\times\vec{S}_3\right)^2+\cdots.
\label{eq:rSDWLandauTheory}
\end{align}
Here $K_5,K_6$ are new symmetry-allowed coefficients.  The important aspect
of this Landau theory is the third-order term which couples the rSDW and rCDW order parameters.  This term is allowed by
symmetry since it is invariant under time-reversal and lattice translation symmetry.

Close to  the transition temperature $T_c$ for the rSDW order, the rSDW-rCDW coupling term can be treated as a
perturbation and we can first solve the rSDW Landau theory. Like the iCDW OPs, the 3Q$+$ and 3Q$-$ classes are related
under time-reversal symmetry so we only need to consider the 3Q and 1Q rSDW classes.
This Landau theory has previously been studied in Ref. \cite{nandkishore_itinerant_2012}.
For chemical potential sufficiently close to the saddle point, the solution to the SDW Landau theory is a uniaxial 3Q rSDW phase
where all three rSDW order parameters have the same magnitude and are
oriented along the same axis, i.e. $\vec{S}_\alpha^* = \vec{s}
n_\alpha$, where $n_\alpha=\pm 1$. If we substitute this solution into Eq.
\eqref{eq:rSDWLandauTheory} and add the terms from Eq. \eqref{eq:rCDWLandauTheory}, we get a free energy that is a function of just the rCDW OPs:
\begin{align}
  &f_\textrm{rCDW}+f_\textrm{rSDW}\big\rvert_{\vec{S}_\alpha=\vec{S}_\alpha^*}
  =\left(\frac{1}{2G_\textrm{rCDW}}+K_1\right)\sum_\alpha N_\alpha^2\notag\\
  &\quad+\frac{K_2}{4}\abs{\vec{s}}^2\left(n_2n_3 N_1+\textrm{cyc. perms.}\right)+K_2N_1N_2N_3\notag\\
  &\quad+K_4\left(\sum_{\alpha}N_\alpha^2\right)^2+(K_3-2K_4)\sum_{\alpha<\beta}N_\alpha^2N_\beta^2.
\end{align}
For $K_3-2K_4<0$, all terms prefer the 3Q$+$ rCDW state, so the system
can develop a rCDW state.

\section{Extensions and experimental implications}
\label{sec:extens-exper-impl}

In this section, we discuss various aspects of the different CDW
phases, and how they may be differentiated experimentally.

\subsection{Real space r/iCDW patterns}
\label{sec:realspaceCDW}
It is interesting to consider the real space patterns of charges and
currents associated with the rCDW and iCDW order parameters.

Our continuum model does not carry any details of the lattice, so we rely on symmetry information provided by DFT calculations~\cite{Teicher}. Importantly, as the saddle point band at the three M points are even under inversion, it can be shown that among the d-orbitals of vanadium atoms, only the $\alpha$-th vanadium atom in the unit cell contributes to the Bloch state at $\veM_\alpha$ (see App. \ref{app:irrep}). Furthermore, the DFT calculation shows that the saddle points consist mostly of d-orbitals of vanadium atoms. Motivated by these observations, we consider nearest neighbor hopping on the Kagom\'e lattice as the minimal tight binding model that should capture the essential physics from fermions near the saddle points. 
For convenience, the lattice coordinate for the three sublattice is expressed as $\ve{r}_\alpha=\ve{R}+\ve{\delta}_\alpha$, where $\ve{R}$ is the coordinate for a unit cell, whose origin is taken at the center of the triangular plaquette with the green sublattice facing to the left in Fig. \ref{fig:lattice}. $\ve{\delta}_1=(-\frac{1}{2\sqrt{3}},0)$, $\ve{\delta}_2=(\frac{1}{4\sqrt{3}},\frac{1}{4})$ and $\ve{\delta}_3=(\frac{1}{4\sqrt{3}},-\frac{1}{4})$. The tight-binding Hamiltonian reads
\begin{align}
\hat{H}_0=t_1\sum_{\langle \ve{r}_{\alpha},\ve{r}_{\alpha}+\ve{e}_\beta \rangle }  d^{\dagger}_{\ve{r}_{\alpha}} d_{\ve{r}_{\alpha}+\ve{e}_\beta}+d^{\dagger}_{\ve{r}_{\alpha}+\ve{e}_\beta}d_{\ve{r}_{\alpha}},
\label{eq:tightbinding}
\end{align}
where $\langle \ve{r}_{\alpha},\ve{r}_{\alpha}+\ve{e}_\beta \rangle$ is the NN bond and $\ve{e}$ is defined such that $\ve{r}_\gamma=\ve{r}_{\alpha}+\ve{e}_\beta$ where $\{\beta, \alpha, \gamma\}$ is a permutation of $\{1,2,3\}$. This gives $\ve{e}_1=\frac{1}{2}\{-1,0\}$, $\ve{e}_2=\frac{1}{2}\{-\frac{\sqrt{3}}{2},\frac{1}{2}\}$ and $\ve{e}_3=\frac{1}{2}\{\frac{\sqrt{3}}{2},\frac{1}{2}\}$.

Due to the one-to-one correspondence between patch label $\veM_\alpha$ and sublattice label $\rm{V}_\alpha$ of vanadium atoms, the real space fermions on vanadium atoms can be expressed as
\begin{align}
d_{\ve{r}_\alpha}= \frac{1}{\sqrt{\mathcal{N}}} \sum_\kv \eu^{\iu \kv \cdot \ve{R}} d_{\alpha,\kv} \approx  \frac{1}{\sqrt{\mathcal{N}}} \sum_{|\qv|<\Lambda} \eu^{\iu \veM_\alpha \cdot \ve{R}} c_{\alpha \qv},
\label{eq:realtofourier}
\end{align}
where $c_{\alpha\qv}$ denotes the saddle point fermion near $\veM_\alpha$ as defined in the continuous model (Eq.~\eqref{eq:noninteractinghamiltonian}). As both rCDW and iCDW order parameters are condensates of inter-patch fermion density operators, the associated real space order must be a \emph{bond order} on the Kagom\'e lattice. For simplicity, we will consider only nearest neighbor (NN) bond order as an example.  

\subsubsection{rCDW pattern}
\begin{figure}[htp]
  \subfigure[]{\includegraphics[width=0.31\linewidth]{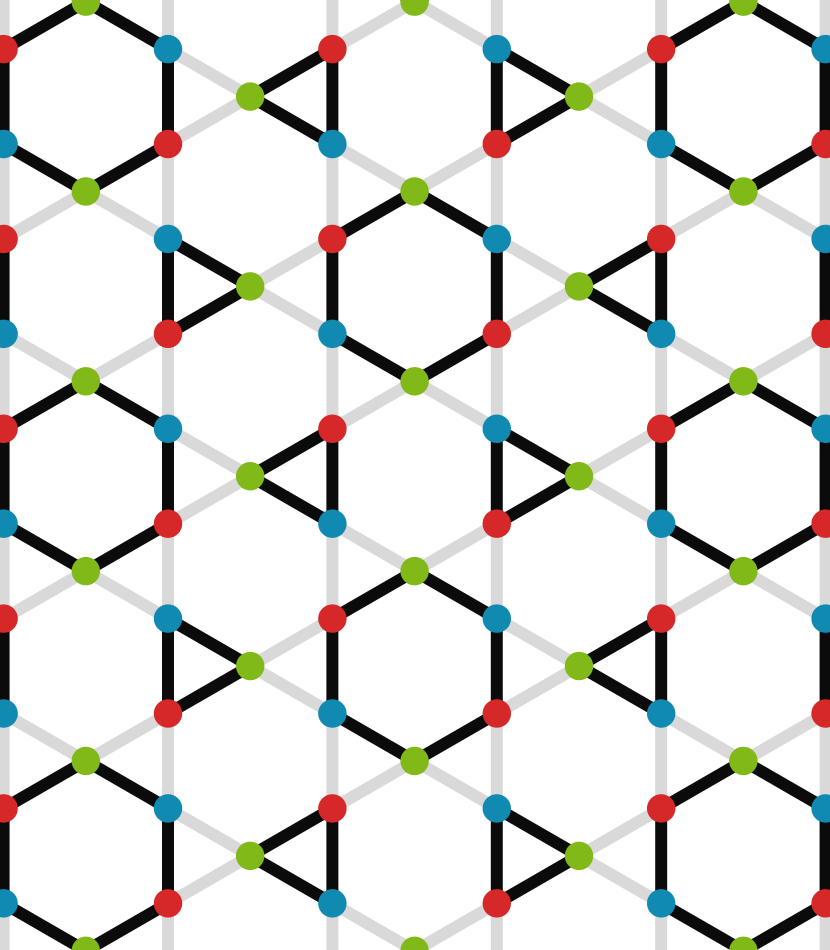}\label{fig:rCDWPatterna}}
  \subfigure[]{\includegraphics[width=0.31\linewidth]{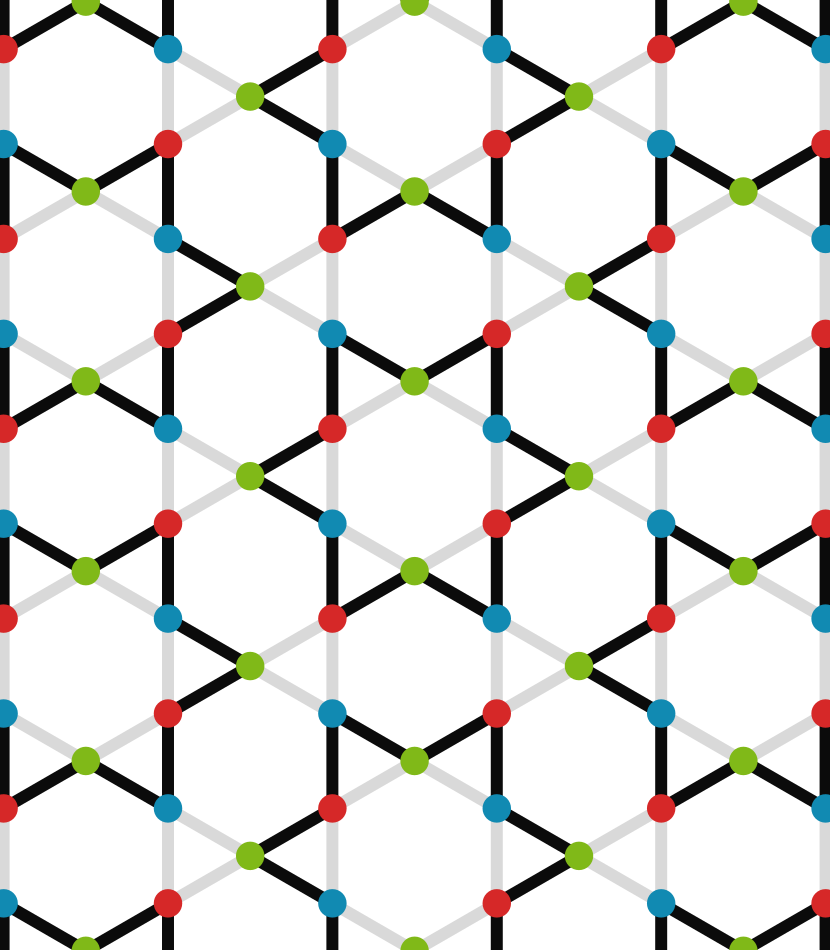}\label{fig:rCDWPatternb}}
  \subfigure[]{\includegraphics[width=0.31\linewidth]{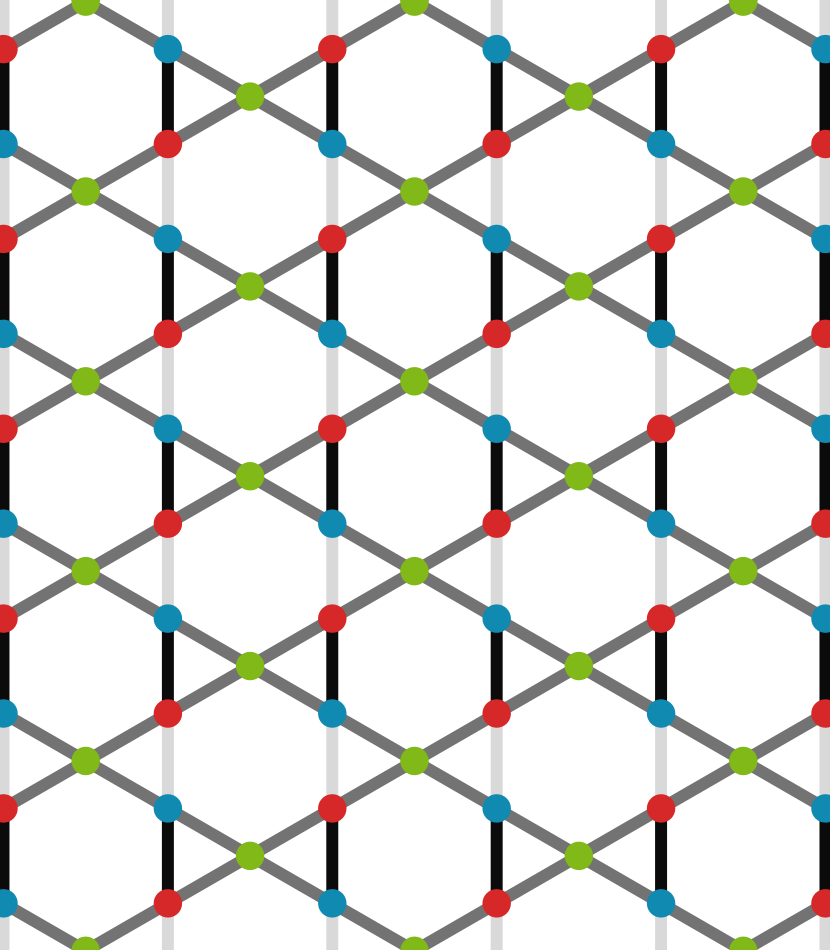}\label{fig:rCDWPatternc}}
  \caption{
  Real space rCDW bond order pattern of the 3Q$\pm$ and 1Q configurations.
  Configurations in the 3Q$\pm$ classes are related by translation. Configurations in the 1Q class are related by
  translation and a three-fold rotation.
  \subref{fig:rCDWPatterna} Bond ordering corresponding to the 3Q$+$ rCDW class. The bond ordering forms hexagonal and triangular plaquettes.
  \subref{fig:rCDWPatternb} Bond ordering corresponding to the 3Q$-$ rCDW class.
  The bond ordering forms a `star-of-David' pattern.
  \subref{fig:rCDWPatternc} Bond ordering pattern corresponding to the 1Q rCDW class. The pattern has alternating bond
  strength along one direction and uniform strength in the other two bond directions.}
  \label{fig:rCDWPattern}
\end{figure}

The rCDW order parameter is time-reversal even, and contribute to bond density modulation as
\begin{align}
\langle \delta \hat{\rho}_{\ve{r}_{\alpha} \ve{r}_{\alpha}+\ve{e}_\beta} \rangle &=\langle \delta \hat{\rho}_{\ve{r}_{\alpha} \ve{r}'_{\gamma}} \rangle=\re \left[\langle d^{\dagger}_{\ve{r}_{\alpha}} d_{\ve{r}_\alpha+\ve{e}_\beta}\rangle\right]\non\\
&=  \frac{N_{\alpha\gamma}}{G_{\rm rCDW}}\cos (\veM_\alpha \cdot \ve{R}-\veM_{\gamma} \cdot \ve{R}') ,
\label{eq:BondCharge}
\end{align}
where $\ve{r}'_{\gamma}=\ve{r}_{\alpha}+\ve{e}_\beta=\ve{R}'+\ve{\delta}_\gamma$ with $\gamma \neq \alpha \neq \beta$.
In Fig.~\ref{fig:rCDWPattern}, the rCDW charge bond density modulation is shown for the 3Q$\pm$ and 1Q rCDW states. 

\subsubsection{iCDW pattern}
\label{sec:icdwpattern}
\begin{figure}[htp]
    \centering
    \subfigure[]{\includegraphics[width=0.4\columnwidth]{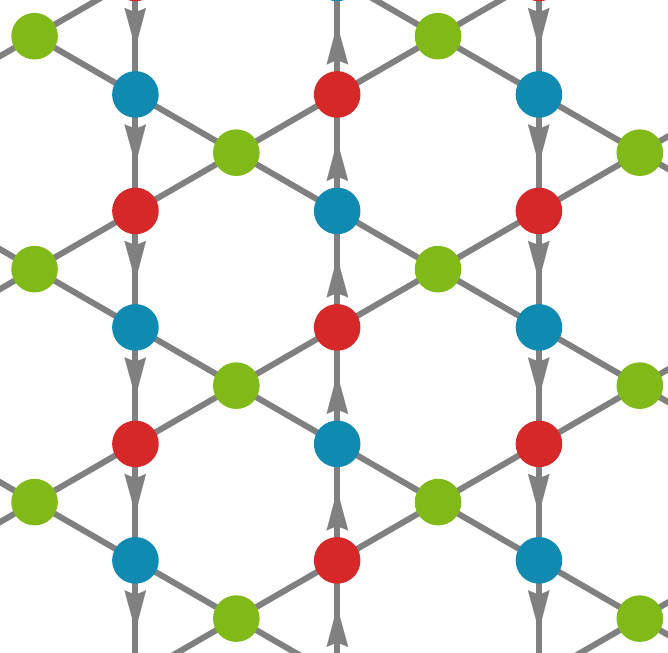}}\quad\quad
    \subfigure[]{\includegraphics[width=0.4\columnwidth]{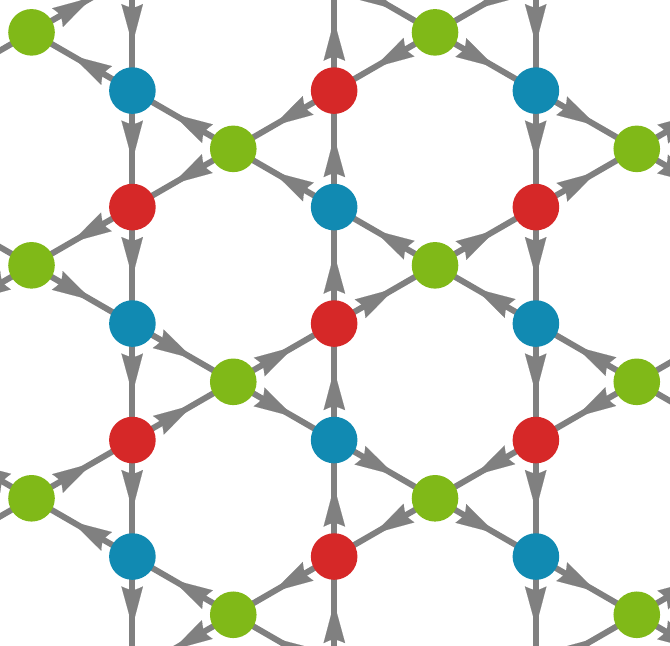}}
    \caption{Real space iCDW (current) bond order pattern of the (a) 1Q and (b) 3Q states. All 3Q configurations are related by translations or time-reversal (changing the sign of the currents).  Configurations in the 1Q class are related by translation (i.e.\ reversing the current direction) and a three-fold rotation. Note that the 1Q state is macroscopically time-reversal symmetric, because it is invariant under the combination of time-reversal and translation. (a) Bond current in the 1Q class, with $\{\phi_{23}, \phi_{31}, \phi_{12}\}=\phi_0 \{1,0,0\}$. (b) Bond current from $\{\phi_{23}, \phi_{31}, \phi_{12}\}=\phi_0\{1,1,1\}$. The sign of $\phi_0$ is taken as negative here. 
    \label{fig:icdw}}
\end{figure}
The iCDW order parameter breaks time-reversal symmetry and corresponds to bond current in real space. 
Note, that the current operator is well defined only when the charge is conserved, which is indeed the case in both the high temperature disordered phase and iCDW phase. As a result, in the equilibrium phase, the current operator must satisfy
\begin{align}
\sum_{\ve{r}_\alpha\rightarrow \ve{r}'_{\alpha'}} \hat{j}_{\ve{r}_\alpha \ve{r}'_{\alpha'}}=0.
\end{align}
By comparing the continuity equation and the equation of motion for charge density, we find the current operator on the NN bonds as (see App.~\ref{app:current}):
\begin{align}
\hat{j}_{\ve{r}_\alpha \ve{r}'_{\alpha'}}&= \hat{j}_{\ve{r}_\alpha \ve{r}_\alpha+\ve{e}_\beta} \delta_{\ve{r}'_{\alpha'}, \ve{r}_\alpha+\ve{e}_\beta} \delta_{\alpha\neq \beta} \non\\
&= \frac{\iu e}{\hbar} \, t_1\left( d^{\dagger}_{\ve{r}_\alpha} d_{\ve{r}_\alpha+\ve{e}_\beta} - d^{\dagger}_{\ve{r}_\alpha+\ve{e}_\beta} d_{\ve{r}_\alpha}\right),
\label{eq:CurrentOP}
\end{align}
where $t_1$ is the NN hopping defined in Eq.~\eqref{eq:tightbinding}.
From Eq.~\eqref{eq:realtofourier}, the bond current expectation value can be expressed in terms of iCDW order parameter as
\begin{align}
&\langle \hat{j}_{\ve{r}_\alpha \ve{r}_\alpha+\ve{e}_\beta} \rangle =\langle \hat{j}_{\ve{r}_\alpha \ve{r}'_{\gamma}} \rangle\non\\
=&  \iu\, e\,t_1\,\eu^{-\iu (\veM_\alpha \cdot \ve{R}-\veM_{\gamma} \cdot \ve{R}')} \left(\frac{1}{\mathcal{N}} \sum_{\qv}\left(\langle c^{\dagger}_{\alpha,\qv}c_{\gamma,\qv}\rangle-\langle c^{\dagger}_{\gamma,\qv}c_{\alpha,\qv}\rangle\right)\right) \non\\
=&\,\frac{-2\,e\,t_1}{G_{\rm iCDW}}\eu^{-\iu (\veM_\alpha \cdot \ve{R}-\veM_{\gamma} \cdot \ve{R}')} \phi_{\alpha\gamma},
\label{eq:BondCurrent}
\end{align}
where again we used $\ve{r}'_{\gamma}=\ve{r}_\alpha+\ve{e}_\beta=\ve{R}'+\ve{\delta}_\gamma$ with $\gamma \neq \alpha \neq \beta$. 
From Eq.~\eqref{eq:BondCurrent}, the real space bond current pattern from $\phi_{23}$ are shown in Fig.~\ref{fig:icdw} (a). The linear combinations of them can form loop current, as an example, we show the bond current pattern for $\{\phi_{23}, \phi_{31}, \phi_{12}\}=\phi_0 \{1,1,1\}$ in Fig.~\ref{fig:icdw} (b). 

The primary order parameter for the iCDW is a loop current.  However, as was discussed in Sec.~\ref{sec:ICDWRCDWMFT},
from the form of the $K_2$ coupling terms in
Eq.~\eqref{eq:iCDWLandauTheory}, a 3Q iCDW will also induce charge
order.  In particular we see that a 3Q iCDW induces either a 3Q$+$ or
3Q$-$ state depending upon the sign of $K_2$.  Notably, the charge order
is quadratic in the iCDW order parameter, $N_\alpha \sim \phi^2$, which
could be detectable near the transition temperature.  

\subsection{Three-dimensional coupling}
\label{sec:three-dimens-coupl}
Here we extend the Landau theory to consider the implications of
coupling of CDW order parameters between nearby layers, adding a layer
index $z=0,1,2\cdots$ numbered beginning from the top layer.  We
assume this coupling is weak, so can be approximated by the leading
terms linear in the order parameters in each layer, and decays rapidly
with the distance between layers.  Hence,
\begin{equation}
  \label{eq:1}
  f_{\perp} = \sum_{z=0}^\infty \sum_{\delta=1}^\infty
  \left( K_{\perp,\delta} N_{\alpha,z} N_{\alpha,z+\delta} + L_{\perp,\delta} \phi_{\alpha,z}\phi_{\alpha,z+\delta}\right).
\end{equation}
We assume that all the
inter-layer interactions are weak compared to the intra-layer terms
in the free energy, so that the form of the order within each layer is
established by the latter, and the intra-layer terms serve to select
particular relative orientations of the different symmetry-breaking
states in nearby layers.    Here we expect
$|K_{\perp,1}|\gg |K_{\perp,\delta>1}|$ and
$|L_{\perp,1}|\gg |L_{\perp,\delta>1}|$ so that terms with $\delta >1$
can be neglected unless they are required to break degeneracies.

\subsubsection{Three-dimensional ground states}
\label{sec:three-dimens-ground}

Now we discuss the resulting three-dimensional ordered structures.
First consider the imaginary CDW, within the 3Q phase.  Within a given
layer, $\phi_\alpha$ may take one of the values
$\phi_\alpha = |\phi_0| (\pm1,\pm1,\pm1)$, where all 8 signs are possible,
and $|\phi_0|$ is fixed by single-layer energetics.  If
$L_{\perp,1}<0$, the minimum energy iCDW order parameter is identical
in all layers, $\phi_{\alpha,z}=\phi_\alpha$.  If instead, $L_{\perp,1}>0$, the
minimum energy configuration is ``antiferromagnetic'',
$\phi_{\alpha,z}=(-1)^{z} \phi_\alpha$.  The two cases above correspond to an
ordering wavevector with the z-component $k_z = 0, \frac{1}{2}$, in
lattice units. Note that $k_z=1/2$ is the wavevector for the
current order, but the induced real CDW order would have $q_z=0$ in
both cases.

Now consider the real CDW, in either the 3Q$+$ or 3Q$-$ states.  Within a single
layer, $N_\alpha$ may take just four values, with $N_\alpha=N_0 n_\alpha$,
with $n_\alpha^2=1$ and $n_1 n_2 n_3 =1$.  For $K_{\perp,1}<0$, we again
obtain a ``ferromagnetic'' state, with $N_{\alpha,z}=N_\alpha$.  For
$K_{\perp,1}>0$, however, the situation is distinct from the iCDW
case, because an overall sign change in $N_\alpha$ is not permitted.
Instead, for a given state $n_{\alpha,z}$, the inter-layer coupling equally
favors $n_{\alpha,z+1}$ in any of the three states not equal to $n_{\alpha,z}$.
For $N$ layers, the total degeneracy of ground states, consider just
the $K_{\perp,1}$ interaction, is $4 \times 3^{N-1}$ , a macroscopic
degeneracy including states with arbitrary wavevectors $k_z$.  We must
therefore consider further neighbor interactions.  The problem can be
mapped to a 4-state Potts model, by defining the four allowed
configurations of $N_\alpha$ in a single layer as $\sigma=1\cdots 4$.  The
interaction energy becomes
\begin{equation}
  \label{eq:2}
   f_{\perp} = \sum_{z=0}^\infty \sum_{\delta=1}^\infty
   K_{\perp,\delta} N_0^2 \left(4 \delta_{\sigma_z,\sigma_{z+\delta}} - 1\right),
 \end{equation}
 which illustrates the $S_4$ permutation symmetry of a Potts model.
 This is a 4-state Potts chain with competing further-neighbor
 interactions, which has a rich statistical mechanics for general
 couplings, similar to that of the ANNNI model, a paradigm for devil's
 staircases, commensurate and incommensurate phases, and transitions
 between them.  Here because each Potts spin represents an entire 2d
 layer, the energies involved are proportional to the area of a layer,
 and hence much larger than $k_B T$.  Therefore, we are interested
 only in the ground states of the Potts chain.  In this limit, the
 ground states are generally commensurate, but can have very large
 unit cells (in the $z$ direction), and the Devil's staircase can
 arise.  We limit our discussion to only the simplest cases, and
 assume $|K_{\perp,\delta}| \gg |K_{\perp,\delta+1}|$ as expected on
 grounds of locality.  

The simplest situation is $K_{\perp,2}<0$, in which case second
neighbor layers prefer to be parallel.  We have then alternating
states in successive layers, $\sigma_z=\sigma_1$ for $z$ even and
$\sigma_z = \sigma_2$ for $z$ odd, with $\sigma_1\neq \sigma_2$.  In
terms of the rCDW vector,, $n_{\alpha,z} = n^{(1)}_\alpha$ for $z$ even and $n_{\alpha,z}=
n^{(2)}_\alpha$ for $z$ odd, such that $\vec{n}^{(1)} \cdot
\vec{n}^{(2)}=-1$.  This corresponds to the wavevector $k_z=1/2$ in
lattice units.  Note that the three-dimensional degeneracy of this
state is $4\times 3=12$.

If $K_{\perp,2}>0$, then we require both nearest neighbor and second
neighbor Potts spins to differ in the ground state.  After choosing
the first layer, there are 3 choices for the second layer, and then 2
choices for the third layer, which must be distinct from the first two
layers.  This implies the smallest possible periodicity of the ground
state is 3.  It may, however, be larger.  Indeed for the fourth
layer, there are still 2 choices remaining, and the ground state is
not determined.  To fix this degeneracy, we may yet consider the third
neighbor coupling.  If $K_{\perp,3}<0$, then we favor a return to the
original state.  The entire configuration becomes determined, with a
periodicity of 3, and a representative sequence in Potts variables
like $\sigma_z = 1,2,3,1,2,3,1,2,3\cdots$.  These configurations
correspond to a ``chiral'' ordering of the layers, consistent with a
3-fold screw axis.  Alternatively, if $K_{\perp,3}>0$, then we obtain
a four layer periodicity, $\sigma_z=1,2,3,4,1,2,3,4,\cdots$.  These
two situations have smaller ordering wavevectors in the $z$ direction,
$k_z=1/3, 1/4$, respectively.  Note that the relatively simple results
quoted here are the result of assuming a strict hierarchy of
interactions, with couplings decaying rapidly in strength with the
separation of layers.  Otherwise much more complex states may arise in
the Potts chain. 

We conclude that a distinct difference between the iCDW and rCDW is
the presence of periodicities larger than 2 in the $c$ direction.
Appearance of such a periodicity (i.e. $k_z=1/3,1/4$) would provide  clear experimental
evidence in favor of the real over the imaginary CDW.   

\subsubsection{Rotational symmetry breaking}
\label{sec:rotat-symm-break}

In the previous discussion, we determined the relative ordering
between layers assuming the order parameter within each layer is
rigid.  Now we consider a higher order effect: the back-influence of
the inter-layer interaction on the order parameter within a single
layer.  In particular, in the case of rCDW order, this leads to a
breaking of $C_3$ rotational symmetry within a given layer.  This can
be understood as follows.  All 4 of the 2d ordered states in the 3Q$+$
or 3Q$-$ states preserve $C_3$ symmetry around the centers of one of the
four hexagons within the quadrupled unit cell, but not around the
other three.  In the states with $k_z>0$, the centers of neighboring
layers are not aligned.  Consequently there is no rotation axis which
preserves all layers.  In the chiral $k_z=1/3$ states, there is
instead a screw axis, which preserves macroscopic $C_3$ symmetry of
the crystal in the bulk.  This symmetry is however broken at the
surface.  In the states with $k_z=1/2,1/4$, there is not even
macroscopic $C_3$ symmetry.

In all cases, if one observes the order
within a single layer, its neighbors will influence its order and
lower the symmetry.  The situation is simplest for the top layer.
Assume the system develops long-range order, and therefore we may
treat the inter-layer interaction in a mean field sense.  We therefore
replace the coupling to the second layer from the top (the strongest
such coupling) by a term of the form
\begin{equation}
  \label{eq:3}
  f_{\perp,0} = K_{\perp,1} \langle N_{\alpha,1}\rangle N_{\alpha,0}.
\end{equation}
This term appears as a ``field'' on the order parameter in the top
layer ($z=0$).  In for example the $k_z=1/2$ phase, we may take
$\langle N_{\alpha,1}\rangle =N_0 (1,1,1)$.  Then this configurations in
the first layer are ``pushed'' away from the $(1,1,1)$ direction.  For
example, if the top layer chooses the $(1,\bar{1},\bar{1})$ state, the
inter-layer coupling will shift it to the form $N_{\alpha,0} =
N_0(1-\delta,-1-\delta,-1-\delta)$, with $0<\delta\ll 1$.  Note that $|N_1|<|N_2|=|N_3|$.
The consequence is that the two dimensional Bragg peaks associated to
density oscillations in the top layer develop two unequal magnitudes.
This is a sign of the rotational symmetry breaking.  Indeed, one can
define a two-dimensional vector $\bm{v}_{z,z'}=-\bm{v}_{z',z}$
\begin{equation}
  \label{eq:4}
  \bm{v}_{z,z'} = \sum_\alpha \langle n_{\alpha,z} n_{\alpha,z'}\rangle \bm{a}_\alpha,
\end{equation}
where $\bm{a}_\alpha$ are the 2d triangular Bravais lattice vectors with
$\bm{a}_3=-\bm{a}_1-\bm{a}_2$.  The vector $\bm{v}_{z,z'}$ is oriented
along one of the three principle directions and selects this axis.

This rotational symmetry breaking effect exists within each layer in
all the 3Q$\pm$ rCDW phases except the uniform $k_z=0$ one.
Rotational symmetry is also broken in the bulk (i.e. in an infinite
system in the $z$ direction) {\em except} when it is restored
macroscopically by an arrangement of layers that constitutes a screw
axis.  The latter occurs only for the $k_z=1/3$ case detailed above.
Rotational symmetry breaking is, however, absent both in individual
layers and in bulk, in the 3Q iCDW states, providing another means to
differentiate the iCDW from the rCDW  experimentally.

\subsection{Critical behavior}
\label{sec:critical-behavior}

We briefly discuss the expected critical behavior at the ordering
temperature for some important cases based on the symmetries and order
parameters.  This question is motivated by the presence of a cubic term in the rCDW Landau theory, which might suggest a first order transition to the CDW state, which to our knowledge is not observed experimentally.  While a full understanding is somewhat involved, we argue below that both thermal fluctuations in two dimensions and three-dimensional coupling stabilize a continuous transition within most scenarios.  

\subsubsection{Two dimensions}
\label{sec:two-dimensions}

If we neglect inter-layer coupling, the problem becomes two
dimensional, and we must be wary of applying a mean field Landau
theory analysis to critical properties, since it is well known that
two dimensional systems have strong thermal fluctuations.

For the case of the 3Q$+$ or 3Q$-$ rCDW, as discussed in
Sec.~\ref{sec:three-dimens-ground}, the ordering can be described by
a four state Potts model.   At mean field level, the $q$-state
Potts models have first order transitions for $q\geq 3$, consistent
with the Landau analysis.  It is well-known, however, that in two
dimensions the $q=3,4$ Potts models in fact have continuous
transitions described by conformal field theory.  In particular, the
$q=4$ Potts model is equivalent to the Ashkin-Teller model, and has
known critical exponents with logarithmic corrections due to marginal
operators (see e.g. Ref.\cite{cardy1980scaling}).

In the case of the 3Q iCDW, the order parameter has a degeneracy of 8
and transforms under full cubic $O_h$ symmetry, and we have seen that
it can be expressed as an O(3) vector with cubic anisotropy.  In two
dimensions, this problem has been analyzed by
Schick\cite{schick1983application}, who concludes that the transition
to the 3Q phase (called ``corner cubic anisotropy'' in this reference)
may be either continuous (and in the Ising universality class,
surprisingly) or first order.

\subsubsection{Three dimensions}
\label{sec:three-dimensions}

The two dimensional critical behavior is valid at best in a regime
close but not too close to the critical point, where a crossover to
three-dimensional behavior must occur.  The nature of the true
three-dimensional critical regime is, however, dependent on the type
of inter-layer couplings, and thereby the three-dimensional order
parameter.  As discussed in Sec.~\ref{sec:three-dimens-ground}, this
is rather complicated for the rCDW problem, and we will not offer a
complete analysis.  The simplest case is the $k_z=0$ one, which occurs
when the inter-layer coupling favors ferromagnetically aligned rCDWs.
Then the symmetry remains unchanged from the four state Potts model.
However, in three dimensions, this model has a first order transition.
Thus in this case a first order transition is predicted.  If instead
one of the $k_z>0$ orderings occur, the situation is less clear, but
it is manifestly {\em not} a Potts model transition.  It is natural to
think that the Landau analysis is more correct in three dimensions.
Based on this reasoning, for the cases $k_z=1/2$ and $k_z=1/4$, a
cubic term in the order parameter is no longer allowed by momentum
conservation.  Thus a continuous transition may be expected.  The case
$k_z=1/3$ seems to allow a cubic term but we will not pursue it
further here.

With three-dimensional coupling, the iCDW order develops either at
$k_z=0$ or $k_z=1/2$.  In both cases, the three-dimensional order
parameter remains an $O(3)$ vector with cubic anisotropy, and the
ground state degeneracy is unchanged from $8$ (this can be seen
because even in the $k_z=1/2$ can a translation by one layer is
equivalent to time-reversal).  Hence the transition should be in the
O(3) cubic universality class, which is known to be continuous, and is
discussed e.g. in Ref.\cite{PhysRevB.61.15136}.

\subsection{Magnetic moment induced by iCDW order}
\label{sec:magn-moment-induc}
The iCDW order can induce both staggered magnetic moment due to the
loop current and uniform magnetic moment. Here, we estimate the
magnitude from each contribution.

First, we note that the bond current obtained in Eq.~\eqref{eq:BondCurrent} is linear in $\phi$, consequently, only the Fourier component at $\veM$ contributes to the bond current, and there is no uniform magnetic moment induced by the bond current. To estimate the staggered magnetic moment, we consider the 3Q iCDW order as an example. As shown in Fig.~\ref{fig:icdw} (b), the bond current forms loop current around both the honeycomb and triangular plaquettes. Treating each plaquette as a current loop, its magnetic moment and the magnetic field it induces can be obtained following the standard magnetostatics~\cite{JacksonEM}. Due to the $2\times 2$ sublattice structure, there are two types of honeycomb plaquettes, with $\vem_{\rm h,1} = I_{\rm e} \hat{\mathcal{A}}_{\rm h}$ for one-quarter of the honeycomb plaquettes, and $\vem_{\rm h,2} = -I_{\rm e} \hat{\mathcal{A}}_{\rm h}/3$ for the rest. Similarly, one-quarter of the triangular plaquettes have $\vem_{\rm t,1} = -I_{\rm e} \hat{\mathcal{A}}_{\rm t }$, and the rest have $\vem_{\rm t,2} = I_{\rm e} \hat{\mathcal{A}}_{\rm t }/3$. Here, $I_{\rm e}=-\frac{e}{\hbar}\frac{2 t_1}{G_{\rm iCDW}} \phi_0$ denotes the magnitude of the electric loop current. $\hat{\mathcal{A}}_{\rm h, t}$ is along $\hat{z}$, such that for $I_{\rm e}>0$, the current flows counter clockwisely. And its magnitude is determined by the area of the honeycomb (h) and triangular (t) plaquette. In unit of Bohr magneton $\mu_B=\frac{e\hbar}{2 m_e}$, we have $\vem \sim I_{\rm e} \hat{\mathcal{A}} = -\frac{2t_1}{G_{\rm iCDW}} \frac{\phi_0}{Ry} \frac{\hat{\mathcal{A}}}{a_b^2} \mu_B$, where $Ry=\frac{m_e e^4}{32\pi^2 \epsilon_0^2 \hbar^2 }\approx 13.6eV$ is the Rydberg energy, $r_b=\frac{4\pi \epsilon_0 \hbar}{m_e e^2} \approx 5.29\times 10^{-11} m$ is the Bohr radius. In $I_{\rm e}$, the factor $\frac{2t_1}{G_{\rm iCDW}}$ is dimensionless and scales with the polarization bubble $\sim \ln^2(t \Lambda^2/T_c) \sim \mathcal{O}(1)$, the order parameter (in unit of energy) at low temperature can be approximated as $\phi_0 \sim k_B T_c$, where $T_c$ is the transition temperature for iCDW order. This gives $\vem_{\rm h,1}\approx -0.047\sgn{(\phi_0)}  \mu_B \hat{z}, \vem_{\rm t,1}\approx 0.008\sgn{(\phi_0)}  \mu_B \hat{z}$, where we take $2t_1/G_{\rm iCDW}=1$. We also note that similar analysis has been done in the literature for cuprates~\cite{Lederer2012} and iron-pnictides~\cite{Klug2018}.

Next, we discuss the uniform magnetization induced by the iCDW order. On the symmetry ground, uniform magnetization requires iCDW order at all three momenta $\veM$ nonzero. Otherwise, the system is invariant under the anti-unitary symmetry composed of time reversal and translation, which forbids any uniform magnetization. In general, the magnetic field couples to the electrons through both the minimal coupling and the Zeeman coupling. Here, we consider the orbital magnetization contribution for the 3Q$\pm$ iCDW order. Following~\cite{Shi2007}, the orbital magnetization in terms of the Bloch wave function reads,
\begin{align}
&\mathcal{M}_{{\rm orb}}^z=\frac{ e}{2\hbar}\sum_{n,\qv}\non\\
&\quad\im\left[\langle \partial_{\qv} {\rm u}_{n,\qv} |\times (\epsilon_{n,\qv} - \hat{H}_{\rm MF}(\qv) )|\partial_{\qv} {\rm u}_{n,\qv} \rangle\right] f_{n,\qv},
\label{eq:MorbDef}
\end{align} 
where $\hat{H}_{\rm MF}(\qv)$ is the mean field Hamiltonian one can infer from Eq.~\eqref{eq:inverseirCDWGF}, $f_{n,\qv}=f(\epsilon_{n,\qv}-\mu)$ is the equilibrium Fermi distribution function. 

For simplicity, we consider the perfect nesting case when the Fermi energy in the disordered phase is at the van-hove point, this gives $1/3$ filling in the patch model. When the 3Q iCDW order is present, the triple degenerate bands at the $\veM$ points are fully gapped, and only the lowest band is filled. Noting that the unit of the summand in Eq.~\eqref{eq:MorbDef} is $\frac{\hbar^2}{2 m_e} $ (coming from the kinetic energy), for a filled band, the summand must be expressed as a function of the only dimensionless parameter in the expression, which reads $\tilde{\epsilon}=\hbar^2 q^2/(2 m_e)$,
 i.e.\ the summand (after averaging over the angular direction in
 $\qv$) must be expressed as $\sim \frac{\hbar^2}{2 m_e}
 \mathcal{F}(\tilde{\epsilon})$. This means physically that for
 electrons within an energy of order the gap $\phi$, the typical
 orbital moment is an order one fraction of a Bohr magneton.

 The orbital magnetization can be expressed as
\begin{align}
\mathcal{M}_{{\rm orb}}^z&=\frac{ e \hbar }{2 m_e} \mathcal{N} \frac{\phi_0}{\hbar^2/(2 m_e a_0^2)} \frac{\sqrt{3}}{2\pi} \left[ \int_0^{\Lambda} \diff \tilde{\epsilon} \mathcal{F}(\tilde{\epsilon})\right]\non\\
&= \mathcal{N}\mu_B \frac{\phi_0}{Ry} \left(\frac{a_0}{a_b}\right)^2\frac{\sqrt{3}}{2\pi} \mathcal{I}_F,
\end{align}
where $\mathcal{N}$ is the number of unit cells, $\mathcal{I}_F=\int_0^{\Lambda}
\diff \tilde{\epsilon} \mathcal{F}(\tilde{\epsilon})$ is the value of
the integral, which is bounded and can be computed numerically as
$\mathcal{I}_F\approx -0.87$. Again approximating $\phi_0\sim k_B
T_c$, the orbital magnetization per unit cell is $\mathcal{M}_{{\rm
    orb}}^z/N \sim -0.017 \sgn(\phi_0)\mu_B$.  

This orbital magnetization is small but the smallness is primarily due
to the small number of electrons within the region near the energy
gap.  As remarked above, for a typical electron within this region,
the orbital moment is a substantial fraction of $\mu_B$.  This
suggests that the iCDW state may be favored by the application of a
magnetic field.

We consider therefore how an rCDW state may be converted into an iCDW
state by such a field, assuming the energy of the iCDW state is not
too much higher than that of the rCDW in zero field.  For simplicity
we consider low temperature, where the above estimate is valid (near
$T_c$, the orbital magnetization will be proportional to $\phi^3$, and
greatly suppressed).  In general, a magnetic field breaks
time-reversal, so will always induce some iCDW component if the rCDW
is present.  We thus take $\Delta_\alpha = N_\alpha + i \phi_\alpha =
|\Delta|e^{i\theta}$ (assuming a 3Q$+$ state), and express the energy in
terms of $|\Delta|$ and $\theta$.  A simplified form for the energy density is
\begin{equation}
  \label{eq:5}
  \mathcal{E} = - m |\Delta|^2 \left(1 + \lambda \cos \theta\right) -
  \mu h m |\Delta|\sin\theta, \end{equation}
where $m$ is the mass scale for the saddle points, $h$ is the magnetic
field (in the $z$ direction), $\mu$ is a typical orbital moment for
the iCDW electrons, and is a fraction of $\mu_B$. The first term is
the condensation energy of the CDW, and the second term is the dipole
energy associated with the orbital magnetization. The parameter
$\lambda>0$ describes the energetic preference for rCDW over iCDW
order in zero field (as well as the preference for 3Q+ over 3Q- order).  When the competition between the two is close,
$\lambda \ll 1$.  Minimizing over $\theta$, we obtain
\begin{equation}
  \label{eq:6}
  \tan\theta = \frac{\mu h}{\lambda |\Delta|}.
\end{equation}
We see that the rCDW smoothly evolves into an iCDW with applied field,
and this occurs on a scale which can be a small fraction of the gap,
if $\lambda \ll 1$.  We note that the energy density in
Eq.~\eqref{eq:5} is simplified, and does not capture topological
physics relevant to the perfect nesting situation when the system is
doped exactly to the saddle point filling.  In this case, the system
is a trivial insulator for $\theta=0$, and a Chern insulator for
$\theta=\pi/2$, and hence there must be a topological transition,
associated gap closing, and non-analyticity of the energy at some
intermediate angle.  However, we expect this distinction to be washed
out away from perfect nesting and for generic filling.

\section{Summary}
\label{sec:summary}

In this paper, we have discussed various mechanisms to induce charge
density wave order in kagom\'e metals, motivated by experiments on
the \AVS\ materials.  We began with a g-ology description based on
a continuum saddle point model, which is quite general to two
dimensional materials with hexagonal symmetry, and was originally
introduced in the context of doped graphene in
Ref.\cite{nandkishore2012chiral}.  We extended the renormalization
group analysis of this problem to the general case, i.e. with both repulsive and attractive bare interactions, and showed that
several distinct density wave and superconducting instabilities are
possible, depending upon parameters. We also showed that the attractive bare interactions may come from coupling of electrons to optical phonons, which induces attraction in certain channels ($g_i$) and lowers the phonon energy. The latter effect may be relevant to the phonon softening observed in ab-initio calculations~\cite{AVSbPhonon}.  Our focus then turned to charge
density wave order at the vectors $\bm{M}_\alpha$, which correspond
both to the locations of the saddle points in the Brillouin zone and
the spanning vectors between them.  They are half reciprocal lattice
vectors and reside at the centers of the zone faces.  We applied a
mean field theory to the real charge density wave (rCDW) states of the
continuum interacting model, and found isotropic 3Q$\pm$ as well as
nematic 1Q rCDW states occur and may be tuned by chemical potential.
We also carried out a Landau theory analysis of the rCDW orders (and a
detailed derivation of the Landau theory coefficients from the continuum
model) which shows that when this is the leading instability, the
above states are the only ones that occur with a high degree of
generality.  Next, we considered the alternate cases of a 3Q imaginary
charge density wave (iCDW) -- actually a state of circulating
currents, i.e. orbital moments -- and a spin density wave (rSDW), which
are competing instabilities with the same wavevectors.  We showed that
both iCDW and rSDW can induce rCDW order, though in both cases the
charge density order is then not the primary order parameter.
Finally, we detailed a number of extensions and implications of the
analysis to derive explicit charge and current patterns, 
three dimensional ordered states, critical properties, and
orbital magnetization of the 3Q iCDW state. 

What are the specific implications of these results for the
\AVS\ materials?  Three distinct compounds, with A=K,Cs,Rb,
have been studied to our knowledge.  All are known to show rCDW order
for which the projection of the ordering wavevector to the 2d kagom\'e
plane is of the 3Q $\bm{M}_\alpha$ type.  A key question is whether
the rCDW is the primary order parameter, or whether it is secondary
and induced, as discussed above, by either an iCDW or rSDW order.
Notably, both the iCDW and rSDW orders break time-reversal symmetry,
and hence should be detectable via their induced local magnetic
moments through e.g. muon spin resonance or neutron scattering.  To
our knowledge, there is no direct evidence of time-reversal breaking
from any measurement in zero magnetic
field (and there is some counter-evidence from muon spin resonance on \KVS\cite{kenney21:_absen_kv3sb}).  This does not definitively exclude
states with very small moments, as indeed may be expected in the iCDW
case (c.f. Sec.~\ref{sec:magn-moment-induc}).  The discussion of
three-dimensional order in Sec.~\ref{sec:three-dimens-ground} provides
some more clear diagnostics.  There, we showed that the iCDW is
expected to display identical induced rCDW order in all layers, i.e. the
z-component of the rCDW ordering wavevector should be zero.  By
contrast, a pure rCDW is consistent with
$k_z=0,\tfrac{1}{2},\tfrac{1}{3},\tfrac{1}{4}$.  Moreover, rCDW states
with $k_z\neq 0$ should show ``nematic'' $C_3$ rotational symmetry
breaking at the surface, due to the three-dimensional coupling, which
again is not expected in the iCDW states.

Several experimental papers are directly related to the above aspects
of the CDW order.  X-ray scattering found Bragg peaks associated with three-dimensional CDW order with $k_z=1/2$ in
\RVS, \CVS\ in Ref.\cite{2021arXiv210309769L}, $k_z=1/4$ in \CVS\ in Ref.\cite{ortiz2021fermi}.  Ref.\cite{2021arXiv210304760L} measured an in-plane shift
of the order in the top two layers of \CVS\ across a step edge using
STM, indicating $k_z>0$.  These observations are consistent with
primary rCDW order.  Ref.\cite{xiang2021nematic} measured a
``nematic'' (i.e. $C_2$ symmetric) dependence of the c-axis
resistivity in \CVS\ on an in-plane magnetic field in the normal (but
CDW) state up to about 60K, 2/3 of the critical temperature.  Such
behavior would be expected below the CDW $T_c$ for any rCDW except the
``screw'' state with $k_z=1/3$.  The STM study in
Ref.~\cite{jiang2020discovery} notes differences in the intensities of
the three charge order wavevector peaks in \KVS.  This reference concludes
that the three intensities are all unequal, and denote this as a
``chiral'' charge order. To our eyes, their Fourier transform STM
data (Fig.3abc of Ref.\cite{jiang2020discovery}) is better described
by two strong and approximately equal intensities and one weaker one,
which is consistent with our theory of inter-layer coupling.  A very recent paper reports the latter type of anisotropy in independent STM measurements of \KVS~\cite{LiRotation}.

While the above evidence seems to favor primary rCDW order, iCDW or
SDW states might be realized in some situations, e.g. as competing
orders that appear at lower temperatures or induced through
applied fields.  Indeed the early observation of large anomalous Hall
effect\cite{yang2019anomalous} in \KVS\ motivated several later
works to promote the possibility of an iCDW (recently similar behavior
was observed in \CVS\cite{Yu2021AHE}) .  As discussed in
Sec.~\ref{sec:magn-moment-induc}, the 3Q iCDW does indeed
macroscopically break time-reversal and is accompanied by a uniform
orbital moment and a large Berry curvature in the vicinity of the
saddle points, which should induce an anomalous Hall effect.  We
discussed how, if iCDW order is only slightly higher in energy than
rCDW, the iCDW can be induced by modest applied fields, which may
provide a possible explanation of Ref.\cite{yang2019anomalous}.  Further
study of the anomalous Hall effect and its correlation with other
measurements is however needed to test this possibility.

While this manuscript was in preparation, several theoretical works
appeared discussing density wave order in these materials. Ref.~\cite{feng2021chiral} evaluated the mean field ground state energies for multiple density wave orders.
Ref.\cite{denner2021analysis} considered a single orbital extended
Hubbard model on the kagom\'e lattice, and within mean field theory
found dominant iCDW order.  Ref.\cite{lin2021complex} discusses a
coupled mean field theory of spin and charge orders and Chern bands in a continuum
saddle point model similar to the one used here and postulates a
complex CDW state in the \AVS\ materials with three unequal
magnitudes of CDW at the three $\bm{M}_\alpha$ wavevectors.  

In this paper, we have focused on density wave order, which is clearly
the dominant instability in the \AVS\ materials.  However, the
superconductivity occurring at lower temperature is of considerable
interest as well.  Since it occurs {\em within} the density wave
ordered state, the understanding of the latter should be important for
developing a theory of superconductivity in this family.

\begin{acknowledgments}
We acknowledge useful discussions with Stephen Wilson, Brenden Ortiz, Sam Teicher, and Ilija Zeljkovic. L.B. is
supported by BES Award DE-SC0020305.  M.Y.\ is supported in part by the Gordon and Betty
Moore Foundation through Grant GBMF8690 to UCSB and by the National Science Foundation under Grant No.\ NSF PHY-1748958.
T.P. was supported by the National Science Foundation through Enabling Quantum Leap: Convergent Accelerated Discovery
Foundries for Quantum Materials Science, Engineering and Information (Q-AMASE-i) award number DMR-1906325.
\end{acknowledgments}

\bibliography{CsV3Sb5}

\appendix
\section*{Appendix}
\section{Irreducible representations at $\veM_\alpha$}\label{app:irrep}
As shown in Fig.~\ref{fig:DFT}, there are two saddle points at $\veM_\alpha$. The little co-group at $\veM_\alpha$ is
$D_\textrm{2h}$, and the irreps of the two saddle points at $\veM_1$ are $\Gamma_1^+$ and $\Gamma_3^+$.  The convention
used here is the one used in Ref. \cite{koster_properties}, and the orientation of the two symmetry axes are shown in Fig.
\ref{fig:lattice}. 

Now, we explain the one-to-one correspondence between the Brillouin zone
patches, $\veM_\alpha$, and the vanadium sites, $V_\alpha$. Consider how Bloch states formed from s-orbitals placed on the vanadium sites ($V_1,V_2,V_3$) transform at $\veM_1$.
A s-orbital Bloch wave on $V_1$ transforms as the irrep $\Gamma_1^+$. On the other hand, s-orbitals placed on $V_2,V_3$ form a
reducible representation that can be decomposed into the $\Gamma_2^-+\Gamma_4^-$ irreps. Notice that the irrep of the
s-orbital on $V_1$ is even under inversion, but the irreps of the s-orbitals on the $V_2,V_3$ sites are odd under
inversion.  Since d-orbitals are even under inversion, this means that Bloch states formed from d-orbitals at $V_1$ must
be even under inversion, but Bloch states formed from d-orbitals on $V_2,V_3$ must be odd under inversion.  The
saddle points at $\veM_1$ are even under inversion; therefore only the d-orbitals on the $V_1$ site contribute to the saddle point at
$\veM_1$.  Using a three-fold rotation, it is easy to see that at $\veM_2, \veM_3$, the contribution to the bands must come from
d-orbitals on the $V_2, V_3$ sites respectively. Hence, there is an one-to-one correspondence between the Brillouin zone
patches, $\veM_\alpha$, and the vanadium sites, $V_\alpha$.  Away from $\veM_1$, d-orbitals on $V_2,V_3$ can start mixing with
the saddle point bands, but we assume these contributions are small sufficiently close to $\veM_1$.

\section{Microscopic mechanism of attractive fermion interaction}
\label{app:elph}

Generally, pure electronic interactions, i.e.\ the screened coulomb repulsion, give rise to repulsive interactions between patch fermions, which favor SDW rather than CDW instabilities. Indeed, the RG phase diagram (Fig.~\ref{fig:RG}) suggests that attraction in the $g^{(0)}_{1}$ or $g^{(0)}_{3}$ channels is likely necessary to stabilize CDW order. Here, we show that the electron-optical phonon interaction may renormalize $g^{(0)}_{1}$ or $g^{(0)}_{3}$, and lead to attraction in these channels. Specifically, we define $g^{(0)}_i=g^{(0)}_{{\rm el},i}+\delta g^{(0)}_i$, where $g^{(0)}_{{\rm el},i}$ is the bare interaction due to screened Coulomb repulsion, and compute the renormalization $\delta g^{(0)}_i$. 

Consider the standard coupling of electron {\em density} with an optical phonon.  Because $g_1$ ($g_3$) transfers spin (charge) between saddle points, these two interactions cannot be generated from coupling to a zone center phonon. Instead, both $\delta g_1^{(0)}$ and $\delta g_3^{(0)}$ require coupling with to a phonon at momentum $\veM$.  Conversely, $\delta g_2^{(0)}$ and $\delta g_4^{(0)}$ are generated from coupling with a zone center phonon. A recent ab-initio calculation and experiment suggest softening of a breathing phonon mode at $\veM$~\cite{AVSbPhonon,UykurOptical2021}.  With these considerations in mind, we focus on the coupling with a phonon at $\veM$. 

To leading ($0^{\rm th}$) order in spatial derivatives, the electron-phonon coupling can be expressed as
\begin{align}
\mathcal{H}_{\rm el-oph}&=\sum_{\qv,\kv, \Gamma,\alpha} g_{\Gamma} u_{\Gamma,-\kv}(\veM_\alpha) \Psi^{\dg}_{\qv+\kv} \Lambda_{\Gamma,\alpha} \Psi_{\qv},
\label{eq:eph}
\end{align}
where $\alpha=1,2,3$ labels the momentum $\veM_\alpha$, $\Gamma$ labels the irreducible representation (irrep) of the little group at $\veM_\alpha$. In this notation, $u_{\Gamma}(\veM_\alpha)$ is the optical phonon mode near $\veM_\alpha$ in irrep $\Gamma$, $\Lambda_{\Gamma, \alpha}$ is obtained such that the fermion bilinear $\Psi^{\dg}_{\qv} \Lambda_{\Gamma,a} \Psi_{\qv} $ transforms as $\Gamma$ irrep at $\veM_\alpha$. Note that in Eq.~\eqref{eq:eph}, the momentum $\kv$ is small, i.e. within the continuum model, and the large momentum transfer $\veM_\alpha$ is treated via the flavor indices of the fermion fields.

Without loss of generality, we consider $\veM_1$, whose little group is ${\rm D}_{2h}=\langle C_2', C_2'', i \rangle$ (see Fig.~\ref{fig:latticeBZ}), where $\langle \cdot \rangle$ indicates the group generated by ``$\cdot$" operations. Here, $C_2'$ and $C_2''$ are two-fold rotation along y-axis and x-axis, respectively. $i$ denotes inversion. Using the transformation of Bloch wave functions, 
\begin{align}
\mathcal{U}_{\mathcal{O}} \psi_{n,\kv}(\ve{r})&=\mathcal{U}_{\mathcal{O}} \eu^{\iu \qv \cdot \ve{r}} {\rm u}_{n,\qv}(\ve{r}) = \eu^{\iu \qv \cdot \mathcal{O}\ve{r}} {\rm u}_{n,\qv}(\mathcal{O}\ve{r})\non\\
&=\eu^{\iu \mathcal{O}^{-1}\qv \cdot \ve{r}} {\rm u}_{n,\mathcal{O}^{-1}\qv}(\ve{r}) = \psi_{n, \mathcal{O}^{-1} \qv}(\ve{r}),
\end{align}
where $\mathcal{O} \in {\rm D}_{2h}$, we find that the fermion bilinear at momentum $\veM_1$ from the patch fermions should be in either the $\Gamma_1^{+}$ or $\Gamma_3^+$ irrep of $\Dh$, with,
\begin{align}
\Lambda_{\Gamma_1^+}=
\begin{pmatrix}
0 & 0 & 0\\
0 & 0 & 1\\
0 & 1 & 0
\end{pmatrix},
\Lambda_{\Gamma_3^+}=
\begin{pmatrix}
0 & 0 & 0\\
0 & 0 & \iu\\
0 & -\iu & 0
\end{pmatrix}.
\end{align}
Similarly, the representation of optical phonon modes at $\veM_1$ in irrep $\Gamma_1^{+}, \Gamma_3^+$ can be identified using~\cite{DresselhausBook},
\begin{align}
\Gamma_{\rm oph}=\Gamma_{\rm equiv}\times \Gamma_{\rm vec},
\end{align} 
where $\Gamma_{\rm equiv}$ is the equivalence representation for the kagom\'e lattice sites at $\ve{M}$, $\Gamma_{\rm vec}$ is the representation for lattice displacement, the same as that for a 3d vector. The $\Gamma_{\rm oph,1}^{+}$ irrep comes from either $\Gamma_{\rm equiv,2}^{-}\times \Gamma_{\rm vec,2}^{-}$ or $\Gamma_{\rm equiv,4}^{-}\times \Gamma_{\rm vec,4}^{-}$ optical phonon modes. While the $\Gamma_{\rm oph,3}^{+}$ irrep comes from either $\Gamma_{\rm equiv,2}^{-}\times \Gamma_{\rm vec,4}^{-}$ or $\Gamma_{\rm equiv,4}^{-}\times \Gamma_{\rm vec,2}^{-}$ optical phonon modes. For example, both the ``star of David" and the ``inverse star of David" breathing phonon mode studied in Ref.~\cite{AVSbPhonon} comes from the $\Gamma_{\rm oph,1}^{+}=\Gamma_{\rm equiv,2}^{-}\times \Gamma_{\rm vec,2}^{-}$ optical phonon mode related by reversing the sign of the lattice displacement field. 

The effective four-fermion interaction renormalized by $\mathcal{H}_{\rm el-oph}$ reads,
\begin{align}
\delta \mathcal{H}_{\rm eff} = \sum_{\qv, \qv', \Gamma, \alpha} \mathcal{V}_{\Gamma} \left(\Psi^{\dg}_{\qv} \Lambda_{\Gamma,\alpha} \Psi_{\qv}\right) \left(\Psi^{\dg}_{\qv'} \Lambda_{\Gamma,\alpha} \Psi_{\qv'}\right),
\end{align}
where $\mathcal{V}_{\Gamma}=g_{\Gamma}^2 \mathcal{D}_{\Gamma}(\veM_1,\Omega_n=0)=-\frac{2g^2_{\Gamma}}{\omega_{\Gamma,\veM_1} }<0$, with $\omega_{\Gamma, \kv}$ the phonon spectrum, $\mathcal{D}_{\Gamma}(\kv, \Omega_n)=\frac{2 \omega_{\Gamma, \kv}}{(i\Omega_n)^2-\omega_{\Gamma, \kv}^2}$ the phonon propagator. This gives finally,
\begin{align}
\delta g_1^{(0)} &=\mathcal{V}_{\Gamma_1^+} +\mathcal{V}_{\Gamma_3^+},\non\\
\delta g_3^{(0)} &=\mathcal{V}_{\Gamma_1^+} -\mathcal{V}_{\Gamma_3^+}.
\end{align}
It is straightforward to see that $\delta g_1^{(0)}<0$ while the sign of $\delta g_3^{(0)}$ depends on the relative strength of $\mathcal{V}_{\Gamma_1^+}$ and $\mathcal{V}_{\Gamma_3^+}$. As a result, all three fixed point solutions may be approached when the electron-phonon interaction is taken into account.

Experimentally, one may test for the significance of electron-phonon coupling by probing the effect on the phonons.  In particular, the coupling $\mathcal{H}_{\rm el-oph}$ also renormalizes the phonon spectrum, and lowers the optical phonon gap. Indeed, the phonon gap for $u_{\Gamma_1^+}, u_{\Gamma_3^+}$ are lowered by $\delta m_{\rm oph}\sim g_\Gamma^2 \Pi_{\rm ph}(\veM)$, which may lead to a Peierls structural transition when the renormalized phonon spectrum becomes gapless at $\veM$.

Does the CDW order observed in the kagom\'e materials come from electronic or structural instabilities? While the full answer to this question requires knowledge of the microscopic details and self-consistent analysis of the electron-phonon coupled system, we note that the four-fermion interaction can be significantly enhanced in the RG flow near the fixed point. Moreover, in the weak coupling limit, as $g^{(0)}_{{\rm el},i}$ is weak, a moderate electron-phonon interaction could alter the sign of the bare fermion interaction, and lead to CDW ordering due to an electronic instability.

\section{Solutions of the rCDW mean-field theory}\label{app:morseDiscussion}
Here, we discuss the possible solutions of the rCDW mean-field theory. The free energy of the system as a function of
the rCDW vector, $\vec{N}=(N_1,N_2,N_3)$, has a tetrahedral point group symmetry, $T_\textrm{d}$. For fixed $\abs{\vec{N}}=N_0$,
rCDW vectors can be categorized into different classes corresponding to the Wyckoff positions of $T_\textrm{d}$. In particular,
there are three distinct rCDW classes that correspond to high symmetry points where the rCDW free energy must have a
critical point on the $\abs{\vec{N}}=N_0$ surface.  They are the 3Q$+$, 3Q$-$, and 1Q classes defined as classes of
rCDW states with the directions,
\begin{center}
\begin{tabular}{l l}
  3Q$+$: & $\{(111),(1\bar{1}\bar{1}),(\bar{1}1\bar{1}),(\bar{1}\bar{1}1)\}$,\\
  3Q$-$: & $\{(\bar{1}\bar{1}\bar{1}),(\bar{1}11),(1\bar{1}1),(11\bar{1})\}$,\\
  1Q: & $\{(\pm100),(0\pm10),(00\pm1)\}$.
\end{tabular}
\end{center}
Configurations in each class are related by $T_d$ symmetry. The critical points of the 3Q$\pm$ class must correspond to
either a maxima or minima. This is because these directions coincide with the three-fold symmetry axes of $T_d$, so the
curvature of the free energy around these points must have the same sign in all directions. The same cannot be said for
the 1Q class which coincides with the two-fold axes of $T_d$, so points in the 1Q directions can either all be saddle
points or maxima/minima. 

For $\abs{\vec{N}}=N_0$, the free energy is a smooth function defined on a sphere. Morse theory tells us that for such a
function, the critical points must satisfy the following constraint, i.e.\ the number of maxima $(n_\textrm{max})$, minima $(n_\textrm{min})$, and saddle points $(n_\textrm{sp})$ must satisfy,
\begin{equation}
  n_\textrm{max}+n_\textrm{min}-n_\textrm{sp}=\chi=2,
  \label{eq:morse}
\end{equation}
where $\chi=2$ is the Euler characteristic of the sphere \cite{milnor_morse_1973}. If we assume that the 3Q$\pm$ and 1Q classes are the only
critical points of the free energy, Eq. \eqref{eq:morse} is satisfied if the 1Q class corresponds to a saddle point
since there are 8 directions in the 3Q$\pm$ classes and 6 directions in the 1Q class. Other solutions to Eq.~\eqref{eq:morse} require additional critical points on the sphere. 
Points on the sphere other than the high symmetry points listed above can be categorized using Wyckoff positions. For $T_d$, there are two
types of low symmetry Wyckoff positions which we label $c,d$ with degeneracies $n_c=12,n_d=24$. It is straightforward to
see that the simplest way to satisfy Eq. \eqref{eq:morse} and include a low symmetry critical point is if there is a saddle point at the Wyckoff positions $c$ or
alternatively, if there is a minimum/maximum at Wyckoff position $c$ and a saddle point at Wyckoff position $d$. In
addition to the cases listed above, pairs of saddle points and maxima/minima can be added to both types of low symmetry
Wyckoff positions while still satisfying Eq. \eqref{eq:morse}. However, we assume that cases with a large degeneracy of
critical points are not physically relevant since they can only come from high-order terms with large powers of $N$ that
are irrelevant at low energies, and correspond to highly oscillatory functions on the sphere. Therefore, given this assumption, the cases where the maxima/minima are only at the
3Q$\pm$ or 1Q positions are expected and the solution to the rCDW mean-field theory should be in of those directions.

A similar analysis can be extended to other order parameters. The iCDW and rSDW order parameters have a cubic point group symmetry, $O_h$, because they are odd under time-reversal symmetry unlike the rCDW order parameters. Therefore, the rCDW states in the 3Q$+$ and 3Q$-$ classes are
related under time-reversal symmetry. Hence, for iCDW and rSDW, the the 3Q$+$ and 3Q$-$ classes are subsets of the
larger 3Q class which is defined as the union of the 3Q$\pm$ classes.

\section{Landau theory coefficients}\label{app:landaucoefficients}
The coefficients, $K_1,K_2,K_3,K_4$, of the Landau theory defined in Eq. \eqref{eq:rCDWLandauTheory} are
\begin{widetext}
\begin{align}
  K_1(T,\mu)=&\frac{1}{2\Lambda^2}\int_{-\Lambda}^\Lambda\frac{\dd^2q}{(2\pi)^2}\frac{f(\varepsilon_2(\vec{q})-\mu)-f(\varepsilon_3(\vec{q})-\mu)}
  {\varepsilon_2(\vec{q})-\varepsilon_3(\vec{q})},
    \label{eq:K1}\\
  K_2(T,\mu)=&-\frac{1}{2\Lambda^2}\int_{-\Lambda}^\Lambda\frac{\dd^2q}{(2\pi)^2}\bigg(
  \frac{f(\varepsilon_1(\vec{q})-\mu)}{(\varepsilon_1(\vec{q})-\varepsilon_2(\vec{q}))(\varepsilon_1(\vec{q})-\varepsilon_3(\vec{q}))}\notag\\
  &\qquad+\frac{f(\varepsilon_2(\vec{q})-\mu)}{(\varepsilon_2(\vec{q})-\varepsilon_1(\vec{q}))(\varepsilon_2(\vec{q})-\varepsilon_3(\vec{q}))}
  +\frac{f(\varepsilon_3(\vec{q})-\mu)}{(\varepsilon_3(\vec{q})-\varepsilon_1(\vec{q}))(\varepsilon_3(\vec{q})-\varepsilon_2(\vec{q}))}\bigg),
  \label{eq:K2}\\
  K_3(T,\mu)=&\frac{1}{24\Lambda^2}\int_{-\Lambda}^\Lambda\frac{\dd^2q}{(2\pi)^2}\bigg(
  \frac{f'(\varepsilon_1(\vec{q})-\mu)}{(\varepsilon_1(\vec{q})-\varepsilon_2(\vec{q}))(\varepsilon_1(\vec{q})-\varepsilon_3(\vec{q}))}\notag\\
  &\qquad+\frac{f'(\varepsilon_2(\vec{q})-\mu)}{(\varepsilon_2(\vec{q})-\varepsilon_1(\vec{q}))(\varepsilon_2(\vec{q})-\varepsilon_3(\vec{q}))}
  +\frac{f'(\varepsilon_3(\vec{q})-\mu)}{(\varepsilon_3(\vec{q})-\varepsilon_1(\vec{q}))(\varepsilon_3(\vec{q})-\varepsilon_2(\vec{q}))}\bigg),
    \label{eq:K3}\\
  K_4(T,\mu)=&-\frac{1}{8\Lambda^2}\int_{-\Lambda}^\Lambda\frac{\dd^2q}{(2\pi)^2}\bigg(
  \frac{f(\varepsilon_2(\vec{q})-\mu)-f(\varepsilon_3(\vec{q})-\mu)}{(\varepsilon_2(\vec{q})-\varepsilon_3(\vec{q}))^3}
  -\frac{f'(\varepsilon_2(\vec{q})-\mu)+f'(\varepsilon_3(\vec{q})-\mu)}{2(\varepsilon_2(\vec{q})-\varepsilon_3(\vec{q}))^2}\bigg),
    \label{eq:K4}
\end{align}
\end{widetext}
where $f(\varepsilon)=1/(\exp(\varepsilon/T)+1)$ is the Fermi-Dirac function. These integrals can be evaluated
asymptotically in the limit $\mu\,,k_BT\ll a\Lambda^2\,,b\Lambda^2$.  If we assume perfect nesting, the patch
dispersions take the form,
\begin{align}
  \varepsilon_1(\vec{q})=&\frac{3t}{4}\left(3q_x^2-q_y^2\right),\\
  \varepsilon_2(\vec{q})=&\frac{3t}{4}2q_y\left(q_y+\sqrt{3}q_x\right),\\
  \varepsilon_3(\vec{q})=&\frac{3t}{4}2q_y\left(q_y-\sqrt{3}q_x\right),
\end{align}
where $t>0$. Given this definition, the asymptotic expressions of $K_2,K_3,K_4$ in the limit $\mu,k_BT\ll t\Lambda^2$
are
\begin{align}
  K_2\approx&-\frac{16}{\pi^2t\Lambda^2k_BT}H_2(\mu/k_BT),\\
  K_3\approx&\frac{8}{3\pi^2\Lambda^2(k_BT)^2}H_3(\mu/k_BT),\\
  K_4\approx&\frac{1}{12\sqrt{3}\pi^2t\Lambda^2(k_BT)^2}H_4(\mu/k_BT)\ln(t\Lambda^2/k_BT),
\end{align}
where $H_2,H_3,H_4$ are the integral functions,
\begin{widetext}
\begin{align}
  H_2(z)=&\int_0^\infty\dd x\int_0^{x/\sqrt{3}}\dd y
  \bigg(\frac{F(\tilde{\varepsilon}_1(x,y)-z)}
  {(\tilde{\varepsilon}_1(x,y)-\tilde{\varepsilon}_2(x,y))(\tilde{\varepsilon}_1(x,y)-\tilde{\varepsilon}_3(x,y))}\notag\\
  &-\frac{F(\tilde{\varepsilon}_2(x,y)-z)}
  {(\tilde{\varepsilon}_1(x,y)-\tilde{\varepsilon}_2(x,y))(\tilde{\varepsilon}_2(x,y)-\tilde{\varepsilon}_3(x,y))}
  +\frac{F(\tilde{\varepsilon}_3(x,y)-z)}
  {(\tilde{\varepsilon}_1(x,y)-\tilde{\varepsilon}_3(x,y))(\tilde{\varepsilon}_2(x,y)-\tilde{\varepsilon}_3(x,y))}\bigg),\\
  H_3(z)=&\int_0^\infty\dd x\int_0^{x/\sqrt{3}}\dd y
  \bigg(\frac{F'(\tilde{\varepsilon}_1(x,y)-z)}
  {(\tilde{\varepsilon}_1(x,y)-\tilde{\varepsilon}_2(x,y))(\tilde{\varepsilon}_1(x,y)-\tilde{\varepsilon}_3(x,y))}\notag\\
  &-\frac{F'(\tilde{\varepsilon}_2(x,y)-z)}
  {(\tilde{\varepsilon}_1(x,y)-\tilde{\varepsilon}_2(x,y))(\tilde{\varepsilon}_2(x,y)-\tilde{\varepsilon}_3(x,y))}
  +\frac{F'(\tilde{\varepsilon}_3(x,y)-z)}
  {(\tilde{\varepsilon}_1(x,y)-\tilde{\varepsilon}_3(x,y))(\tilde{\varepsilon}_2(x,y)-\tilde{\varepsilon}_3(x,y))}\bigg),\\
  H_4(z)=&\int_0^\infty\dd v\left(\frac{-F(v-z)+F(-v-z)}{v^3}+\frac{F'(v-z)+F'(-v-z)}{z^2}\right).
\end{align}
\end{widetext}
In the equations above, $F(x)=1/(e^x+1)$, and $\tilde{\varepsilon}_1(x,y)=3x^2-y^2\,,
\tilde{\varepsilon}_2(x,y)=2y(y+\sqrt{3}x)\,, \tilde{\varepsilon}_3(x,y)=2y(y-\sqrt{3}x).$ The expression holds for
arbitrary ratios of $\mu/k_BT$.  The values of the $H_2,H_3,H_4$ when $\mu\ll k_BT$ are
\begin{align}
  H_2(0)\approx&0.039571,\\
  H_3(0)\approx&0.014216,\\
  H_4(0)\approx&0.213139.
\end{align}
$H_2,H_3,H_4$ each change sign once at the following values:
\begin{align}
  z_{0,2}\approx&2.14268,\\
  z_{0,3}\approx&4.05161,\\
  z_{0,4}\approx&1.91067.
\end{align}
The expression for $K_1$ can be evaluated asymptotically when $\mu\ll k_BT$:
\begin{equation}
  K_1\approx-\frac{1}{6\sqrt{3}\pi^2 t\Lambda^2}\left(\ln(t\Lambda^2/k_BT)^2+2\ln 3\ln(t\Lambda^2/k_BT)\right).
\end{equation}

For $\mu\ll T$, $K_3-2K_4$ which determines whether the quartic term of the rCDW Landau theory prefers a 3Q$+$ or 3Q$-$ configuration
is
\begin{equation}
    K_3-2K_4\approx\frac{1}{\pi^2t\Lambda^2(k_BT)^2}\left(\frac{8H_3(0)}{3}-\frac{H_4(0)}{12\sqrt{3}}\ln(t\Lambda^2/k_BT)\right).
\end{equation}
Therefore, $K_3-2K_4<0$ when
\begin{equation}
    \frac{\mu}{t\Lambda^2}\ll\frac{k_BT}{t\Lambda^2}<\exp(-32\sqrt{3}\frac{H_3(0)}{H_4(0)})\approx0.0248.
\end{equation}

\section{Derivation of Eq.~\eqref{eq:CurrentOP}: the current operator}
\label{app:current}
Here, we show more details to obtain the bond current operator, which is defined from the continuity equation of charge density. 
If we define the charge density at site $\ve{r}_\alpha$ as $\hat{n}_{\ve{r}_\alpha}=-e\,d^{\dagger}_{\ve{r}_\alpha} d_{\ve{r}_\alpha}$, the lattice version of the continuity equation reads,
\begin{align}
\frac{\diff \hat{n}_{\ve{r}_\alpha}}{\diff \rm t} + \sum_{\ve{r}_\alpha\rightarrow \ve{r}'_{\alpha'}} \hat{j}_{\ve{r}_\alpha \ve{r}'_{\alpha'}}=0.
\end{align}
The expression for $\hat{j}$ can be obtained from the equation of motion of $ \hat{n}_{\ve{r}_\alpha}$ which reads,
\begin{align}
\frac{\diff \hat{n}_{\ve{r}_\alpha}}{\diff \rm t}=-\frac{\iu }{\hbar} \left[\hat{n}_{\ve{r}_\alpha}, \hat{H}\right] =-\frac{\iu }{\hbar}\left[\hat{n}_{\ve{r}_\alpha}, \hat{H}_0\right]+...
\end{align}
where ``..." denotes contribution from $\left[\hat{n}_{\ve{r}_\alpha}, \hat{H}_1\right]$, which only possibly contribute at $\mathcal{O}(\phi^2)$ to the bond-current order parameter $\langle \hat{j} \rangle$. On the other hand, as only odd power in $\phi$ term breaks time reversal symmetry, we conclude that $\hat{H}_1$ should not contribute to the bond-current order parameter at the mean field level. Hereafter, only the contribution from $\hat{H}_0$ is considered. The NN bond order requires NN hopping term in $\hat{H}_0$. Noting,
 \begin{align}
\left[\hat{n}_{\ve{r}_\alpha}, \hat{H}_0\right] &= \left[\hat{n}_{\ve{r}_\alpha}, t_1\sum_{\langle \ve{r}'_{\alpha'},\ve{r}'_{\alpha'}+\ve{e}_\beta \rangle }  d^{\dagger}_{\ve{r}'_{\alpha'}} d_{\ve{r}'_{\alpha'}+\ve{e}_\beta}+d^{\dagger}_{\ve{r}'_{\alpha'}+\ve{e}_\beta}d_{\ve{r}'_{\alpha'}} \right]\non\\
&= \left[\hat{n}_{\ve{r}_\alpha}, t_1\sum_{\alpha\neq \beta }  d^{\dagger}_{\ve{r}_\alpha} d_{\ve{r}_\alpha+\ve{e}_\beta}+d^{\dagger}_{\ve{r}_\alpha+\ve{e}_\beta}d_{\ve{r}_\alpha}\right]\non\\
&= -e\,t_1\sum_{\alpha\neq \beta } \left( d^{\dagger}_{\ve{r}_\alpha} d_{\ve{r}_\alpha+\ve{e}_\beta} - d^{\dagger}_{\ve{r}_\alpha+\ve{e}_\beta} d_{\ve{r}_\alpha}\right),
\end{align}
where the second line comes from the fact that the commutator is nonzero only for bond satisfying $\ve{r}_\alpha=\ve{r}'_{\alpha'}$ or $\ve{r}_\alpha=\ve{r}'_{\alpha'}+\ve{e}_\beta$, we find Eq.~\eqref{eq:CurrentOP} in the main text.

\end{document}